%% file: manu.tex
\documentclass[tighten,twocolumn,appendixfloats,apj]{likeapj}
\pdfoutput=1

\usepackage{cfr-lm}

\usepackage{graphicx,bm,mathptmx,amsmath,float}

\hypersetup{linkcolor=blue,citecolor=blue,filecolor=cyan,urlcolor=blue,bookmarksnumbered=true,bookmarksopen=true}
\usepackage[all]{hypcap}
\usepackage{natbib}
\usepackage{graphicx}
\usepackage{epsfig}
\usepackage{array}
\usepackage{multirow}

\newcolumntype{C}[1]{>{\centering}m{#1}}

\include{newcomm}

\newcommand{\mrw}[1]{{\multirow{2}{*}{#1}}}

\received{2019 January 31}
\revised{2019 August 8}
\accepted{2019 September 5}
\published{2019 October}
\shorttitle{None}
\shortauthors{Tzanavaris et al.}

\begin{document}

\title{\hspace{-1cm}Broad Band X-ray Constraints on the Accreting Black Hole in Quasar \fourc} 

\author{P.~Tzanavaris}%\altaffiliation{Not-A Fellow}
\affil{University of Maryland, Baltimore County, 1000
  Hilltop Circle, Baltimore, MD 21250, USA}
\affil{Laboratory for X-ray Astrophysics, NASA/Goddard
  Spaceflight Center, Mail Code 662, Greenbelt, MD 20771, USA}

\author{T.~Yaqoob}
\affil{University of Maryland, Baltimore County, 1000
  Hilltop Circle, Baltimore, MD 21250, USA}
\affil{Laboratory for X-ray Astrophysics, NASA/Goddard
  Spaceflight Center, Mail Code 662, Greenbelt, MD 20771, USA}
\affil{Department of Physics and Astronomy, The Johns
  Hopkins University, Baltimore, MD 21218, USA}

\author{S.~LaMassa}
\affil{Space Telescope Science Institute, 3700 San
  Martin Drive, Baltimore, MD 21218, USA}

\author{M.~Yukita}
\affil{Laboratory for X-ray Astrophysics, NASA/Goddard
  Spaceflight Center, Mail Code 662, Greenbelt, MD 20771, USA}
\affil{Department of Physics and Astronomy, The Johns
  Hopkins University, Baltimore, MD 21218, USA}

\author{A.~Ptak}
\affil{Laboratory for X-ray Astrophysics, NASA/Goddard
  Spaceflight Center, Mail Code 662, Greenbelt, MD 20771, USA}
\affil{Department of Physics and Astronomy, The Johns
  Hopkins University, Baltimore, MD 21218, USA}

\begin{abstract}

  \x\ data for quasar \fourc\ have previously been modeled with a broad \feka\ emission line and reflection continuum originating in the inner part of the accretion disk around the central supermassive black hole (SMBH), i.e.~the strong gravity regime. We modeled broadband \x\ spectra from \suzaku\ and \nustar\ with \myt, self-consistently accounting for \feka\ line emission, as well as direct and reflected continuum emission, from finite column density matter. A narrow \feka\ emission line originating in an \x\ reprocessor with solar Fe abundance far from the central SMBH is sufficient to produce excellent fits for all spectra.
  For the first time, we are able to measure the global, out of the line-of-sight column density to be in the range $\sim$1.5 to $\sim$\ten{2.9}{24} \cunits, i.e.~in the Compton thick regime, while the line-of-sight column density is Compton thin in all observations. The \feka\ emission line is unresolved in all but one observations. The Compton scattered continuum from distant matter removes the need for relativistic broadening of the \feka\ emission line, which is required for SMBH spin measurements.
  %Our tests for a broad line component show that it is not supported by the data,with the possible exception of one case that can alternatively be modeled with a relativistic model with a broken power law emissivity.
    The resolved line observation can alternatively be modeled with a relativistic model but we do not find evidence for a truncated accretion disk model.
  We conclude that the \x\ emission in these \fourc\ data is unlikely to originate in the inner accretion disk region and thus cannot be used to measure SMBH spin.
\end{abstract}


\section{Introduction}\label{sec-intro}
All galaxies with bulges are thought to harbor supermassive black
holes (SMBHs, $\sim$$10^6 - 10^9$ \msun) in their centers
\citep[see][for a review]{graham2016}. The resulting strong gravity in
the nuclear region leads to gravitational collapse of surrounding
material via disk accretion onto the nuclear SMBH, releasing large
amounts of energy. In turn, a population of energetic, hot electrons,
likely residing in a diffuse, hot corona, are thought to induce
thermal inverse-Compton upscattering of optical/UV photons from the
accretion disk, producing a primary \x\ power-law continuum, which is
one of the key signatures of Active Galactic Nuclei (AGNs). This
continuum may then irradiate the accretion disk, or gaseous structures
such as clumpy clouds or a torus that are more distant from the
central SMBH.  The interaction of the primary \x\ continuum with
optically thick, cold, neutral and/or mildly ionized material can give rise
to a series of fluorescent emission lines, and associated
Compton-scattered, or ``reflected'' continua.  Due to a combination
of cosmic abundance and fluorescent yield, most prominent is the
\feka\ emission line at 6.4~keV ($\sim$1.94\AA) and its associated reflection
continuum \citep{george1991}.
%\fetwosix\ ($\sim 6.97$~keV).

According to the ``AGN unification scheme''
\citep{antonucci1993,urry1995}, galaxies with accreting SMBHs (Active
Galactic Nuclei, AGNs) are broadly classified as Type 1 or 2. The
distinction is based on viewing angle, so that Type 1s are viewed
face-on, and Type2s edge-on; intermediate types are also
possible. As a result, Type 1s are more likely to afford
a direct view of the strong gravity region as their
line-of-sight to the central engine is less likely to be obscured.

The \feka\ emission line
comes in two types, often referred to simply as {\it narrow} and {\it
  broad} line.  The ``narrow core'' component of the line
(FWHM~$\lesssim 2,000$ \kmps) is detected in the great majority of
both Type 1 and Type 2 AGNs with luminosities \lxht~$<10^{45}$
\lunits\ \citep{yaqoob2004,nandra2006,shu2010,shu2011,fukazawa2011,ricci2014}.
An additional, relativistically broadened component (FWHM several
thousands to tens of thousands \kmps) is thought to
be widespread and found in {\it at least} $\sim$36\%\ of AGNs
(\citealt{delacalleperez2010}; see also, e.g.,
\citealt{porquet2004,jimenez-bailon2005,guainazzi2006,nandra2007,brenneman2009,patrick2012,liuz2015,mantovani2016,baronchelli2018}).

The smaller FWHM of the narrow line suggests that the reprocessing of
the \x\ continuum is occurring in distant matter at hundreds to thousands of
gravitational radii from the SMBH and its immediate
environment. Measurement of the properties of the narrow line and its
associated reflection continuum then provides unique constraints to
the physical properties of the larger scale structure associated with
the central engine.  In contrast, the broad line should have its
origin closer to the SMBH, implying that reprocessing of the intrinsic
X-ray continuum is taking place at the accretion disk itself. In this
case the width of the line is a combination of Doppler and general
relativistic effects, which lead to line broadening or ``blurring'',
with the contributions becoming progressively stronger as one
approaches, and up to, the innermost stable circular orbit (ISCO), the
closest limit to the black hole where the disk remains
geometrically-thin, optically-thick, and radiatively efficient
\citep[for reviews
  see][]{fabian1989,reynolds2003,miller2007,reynolds2016}. Because
the ISCO location depends directly on black-hole spin, the latter
affects the profile of the observed relativistically broadened
line. In practice, the effect of black-hole spin on the line profile
is not definitive, as it is degenerate with the unknown radial line
emissivity profile and Fe abundance. Moreover, the effects of
black-hole spin manifest themselves at the extreme energy limits of the line
profile, i.e.~regimes that are most sensitive to correct deconvolution of
the underlying continuum (which must be deduced from spectral fitting
simultaneously with the line profile). Nevertheless, these effects
have been modeled in detail, forming the basis of the X-ray reflection
method of spin determination. As a result, there is a large number of
black hole spin measurements in the literature both for X-ray binaries
\citep[][and references therein]{middleton2015} and AGNs \citep[][and
  references therein]{brenneman2013}.  Constraining black hole spin
represents one of the holy grails of astrophysical science, with
far-reaching implications on our understanding of black holes and
their impact on their surrounding environment.  It is an essential
step in any Kerr-metric based test of strong-field General Relativity.
In addition it can provide insight into the mechanism driving nuclear
jets, such as the \citet{blandford1977} paradigm of rotational energy
extraction from a central Kerr (i.e.~rotating) black hole, which
directly affects galaxy environment and by extension galaxy evolution.

However, the results for spin measurements via the reflection method
to date span the allowed spin values, even for the same object and
data. At the same time, best-fitting results for some of the most well
known broad line AGNs often require highly supersolar Fe abundances,
such as up to $\sim 4\times$solar for
\mstf\ \citep[e.g.][]{chiang2011}, $\sim 4.6$ or $\sim 4\times$solar
for NCG~3783 (\citealt{brenneman2011} and \citealt{reynolds2012}), or
even $>8.4\times$solar for Fairall~9 \citep{lohfink2012}.
\citet{patrick2012} noted the degeneracy between spin and Fe
abundance, with high spin magnitudes requiring supersolar
abundances. In addition, it has now been shown that some AGNs
previously well-known for harboring a broad \feka\ emission line can
be modeled exclusively with a narrow line from distant matter and the
reflection continuum associated with it, using only solar Fe abundance
(e.g.~\citealt{yaqoob2016} for Fairall~9; \citealt{murphy2014} for
MCG$+8$$-$11$-$11).
In such cases, since a broad \feka\ emission line is not required in the model, black-hole spin determination from the X-ray spectrum is not possible.

The quasar \fourc\ is one of the nearest ($z=0.104$) FRII powerful
broad-line radio galaxies \citep{riley1989}.
\citet{woo2002} measured a bolometric
luminosity \lbol~$\sim 2\times 10^{46}$ \lunits\ and estimated a
black hole mass $\sim4\times 10^9$ \msun. In the optical,
permitted lines have been reported \citep[$\sim8000 - 11,000$
  \kmps][]{riley1989,corbin1997,brinkmann1998,robinson1999} that
are very broad, as far as optical lines are concerned.

In \fourc\ most previous works on the \x\ spectrum, using a variety of telescopes, instruments, and models reported a strong, broad \feka\ line with a width equivalent to a Gaussian $\sigma$ of up to $\sim 600$~eV, corresponding to a velocity FWHM of $\sim 0.22 c$, where $c$ is the speed of light. An equivalent width (EW) of up to $\sim 300$~eV was reported for the broad \feka\ line, a value that is not atypical of broad \feka\ lines reported in other AGNs. Only sometimes was an accompanying narrow component also reported.
The earliest
modeling results with the {\it Advanced Satellite for
Cosmology and Astrophysics} (\asca) included either a narrow only
\citep{brinkmann1998} or a broad only
\citep{sambruna1999,hasenkopf2002} line. \citet{hasenkopf2002}
reported a broad only line for the same \asca, and also {\it BeppoSAX}
data.  In more recent work, a broad line was reported by
\citet{ballantyne2005,ballantyne2005b,larsson2008,patrick2012,tombesi2014,gofford2013,digesu2016,lohfink2017,bhatta2018}. In
contrast, \citet{noda2013} reported a narrow line only. Among these
results supersolar abundances were reported by \citet[][up to
  4.8$\times$solar]{larsson2008}, while \citet{gofford2013} reported
Fe {\it under-}abundances of $\sim$0.3$\times$solar. \citet{patrick2012} and \citet{gofford2013} also
reported fitting a narrow line component.
Further, four
works estimated or assumed SMBH spin values for this
and \citet{bhatta2018}
assumed a maximally spinning black hole (spin value of 1),
\citet{lohfink2017} reported constraining spin values to $>$0.5.
Further details on
previous work are given in Appendix~\ref{app-prev}.

\fourc\ is an excellent test case for broad \feka\ line modeling.
Among AGNs for which a broad line has been claimed to be prominent, it
has a high luminosity and low redshift, thus occupying a
region of parameter space that has not been thoroughly investigated.
In addition, its \x\ spectrum is not particularly complicated,
representing a relatively straightforward case for \x\ spectral
modeling.

\renewcommand{\arraystretch}{0.7}
\input{tab-suznu}

\renewcommand{\arraystretch}{1}

Note that studies of AGNs have generally only measured the
line-of-sight equivalent hydrogen column density
(labeled \nhz\ in this paper, \scr{sec-mytorus}).
However, a model such as \myt\ \citep{murphy2009,yaqoob2012}
can measure the line-of-sight as well as the {\it global}, out of the
line-of-sight column density
(labeled \nhs\ in this paper, \scr{sec-mytorus}).
This is crucial because a key ingredient for population synthesis
modeling of the Cosmic X-ray Background (CXB) spectrum is the number
density of ``Compton-thick'' (highly obscured, equivalent hydrogen
column density ~$\gtrsim$$10^{24}$~\cunits) versus ``Compton-thin''
AGNs. However, traditional analyses that use the ``Compton-thick
fraction'' as a parameter do not distinguish between sources that are
Compton-thick {\it only} in the line-of-sight and those that are {\it
  globally} Compton-thick; yet such a distinction affects the observed
X-ray spectrum. Measuring the separate column densities is then
critical, as one of them may be Compton-thick and the other one
Compton-thin \citep{lamassa2014,yaqoob2015}.

\input{fig-NHplot}

\vspace{0.2cm}
In this paper we test the hypothesis that the \suzaku\ and
\nustar\ spectra of \fourc\ can be fitted only with a narrow
\feka\ line emission component and associated reflected continuum.
The implication of this would be that, since the narrow
\feka\ emission line does not originate in the strong gravity regime,
it would not be possible to measure black hole spin.  We use \myt,
which self-consistently models the \feka\ line emission doublet
together with its associated reflection continuum and solar Fe
abundance.
Unlike previous work, this also allows us to constrain
the equivalent hydrogen column density both into and out of the line-of-sight independently.
\myt\ models
the Compton reflection continuum from matter with a finite column
density, which can masquerade as a broad component of the
\feka\ line, and more so than the usual reflection from infinite column
density disk models.

\vspace{0.2cm}
The structure of the paper is as follows. In \scr{sec-obs} we describe
the observations and data reduction. Spectral modeling methodology is
discussed in \scr{sec-modeling}. Results are presented and discussed
in \scr{sec-results}. Finally, a summary and conclusions are presented
in \scr{sec-summ}.  We use a concordance cosmology, namely $H_0 = 70$
\kmps~Mpc\up{-1}, $\Omega_{\Lambda} = 0.73$, $\Omega_M = 0.27$
throughout.

\input{fig-NHS-Cpintoxis}
  \renewcommand{\arraystretch}{0.7} 
\input{tab-nh}

  \renewcommand{\arraystretch}{1} 

%\vspace{-.5cm}
\section{Observations and Data Reduction}\label{sec-obs}
In this paper we use two \suzaku\ and four \nustar\ archival
observations.  Details of the observations are given in
\tr{tab-suznu}, including references from the literature that report
results of previous analyses of these observations. As can be seen
from Column 6, both \suzaku\ and \nustar\ have broadband \x\ coverage,
which is critical for modeling the reflection continuum well.

\vspace{-.4cm}
\subsection{\suzaku}
We study two archival observations of \fourc\ carried out by the joint
Japan/US \x\ astronomy satellite, \suzaku\ \citep{mitsuda2007}. For
brevity, we refer to the \suzaku\ ObsIDs as 702 and 706.

\suzaku\ had four \x\ Imaging spectrometers
\citep[XIS,][]{koyama2007} and a collimated Hard \x\ Detector
\citep[HXD,][]{takahashi2007}.  Each XIS consisted of four CCD detectors
with a field of view $17.8\times17.8$~arcmin\up{2}. Of the three front-side
illuminated (FI) CCDs, (XIS0, 2, and 3) XIS2 had ceased to operate
prior to the observations studied here.  We thus used FI CCDs XIS0 and
3, as well as the back-side illuminated (BI) XIS1.  The operational
bandpass is (0.2--)~0.4--12~keV for (BI) FI. However, the useful
bandpass depends on the signal-to-noise ratio of the
background-subtracted source data since the effective area strongly
diminishes at the ends of the operational bandpass.  The HXD consisted
of two non-imaging instruments (the PIN and GSO) with combined
bandpass $\sim$10--600~keV. Both instruments are background limited,
with the GSO having the smaller effective area. We only used the PIN
data as the GSO data did not provide a reliable spectrum.
ObsID 702 was acquired at the ``HXD nominal'', and 706 at the
``XIS nominal'' position.
%The observations
%were acquired at the XIS nominal position, at the expense of
%$\sim10\%$ lower HXD effective area.

\input{fig-NuSTAR}
\input{fig-702-xis}
\input{fig-702-xispin}
\input{fig-706-xis}
\input{fig-706-xispin}

The principal data selection and screening criteria for the XIS were
the selection of only ASCA grades 0, 2, 3, 4, and 6, the removal of
flickering pixels with the FTOOL {\tt cleansis}, and exclusion of data
taken during satellite passages through the South Atlantic Anomaly (SAA), as
well as for time intervals less than 256~s after passages through the
SAA, using the T\_SAA\_HXD house-keeping parameter. Data were also
rejected for Earth elevation angles (ELV) less than 5\degr, Earth
day-time elevation angles (DYE\_ELV) less than 20\degr, and values of
the magnetic cut-off rigidity less than 6 GeV/c\up{2}.  Residual
uncertainties in the XIS energy scale are of the order of 0.2\%\ or
less (or $\sim$13~eV at 6.4~keV).  The cleaning and data selection
resulted in the net exposure times shown in \tr{tab-suznu}.

We extracted XIS source spectra in a circular extraction region with a
radius of 3.5 arcmin.  We constructed background XIS spectra
from off-source areas of the detector, after removing a circular
region with a radius of 4.5 arcmin centered on the source, as well as
the calibration sources (using rectangular masks). The
background-subtraction method for the HXD/PIN used the files
{\tt ae702057010\_hxd\_pinbgd.evt}, {\tt ae706028010\_hxd\_pinbgd.evt},
corresponding to the ``tuned'' version of the background model.

Spectral response matrix files (RMFs) and telescope effective area
files (ARFs) for the XIS data were made using the mission-specific
FTOOLS XISRMFGEN and XISSIMARFGEN, respectively. The XIS spectra from
XIS0, XIS1, and XIS3 were combined into a single spectrum for spectral
fitting. The three RMFs and ARFs were all combined, using the
appropriate weighting (according to the count rates and exposure times
for each XIS), into a single response file for the combined XIS
background-subtracted spectrum. For the HXD/PIN spectrum, the supplied
spectral response matrices appropriate for the times and nominal
pointing mode of the observations ({\tt ae\_hxd\_pinhxnome4\_20080129.rsp
} for 702 and {\tt ae\_hxd\_pinxinome11\_20110601.rsp} for 706) were used
for spectral fitting.

%This is what we discussed, not like yaqoob2016
We determined useful energy bandpasses for the spectrum from each
instrument by first assessing background subtraction systematics.  For
XIS we used spectra with a uniform binning, with 30 eV bin width, and
found that in the 0.3--9.5~keV and 0.3--10.0~keV bands, for
observations 702 and 706 respectively, there were $>$20 counts per bin
for the unscaled background, total source, and background-subtracted
source. In addition, in these regions, the background counts, scaled
by the relative source vs.~background region areas, were $<$50\%\ of
the background-subtracted source counts. Since the counts per bin were
$>$20 in the stated energy bands, we were able to use the $\chi^{2}$
statistic for spectral fitting.
Note that we did not group spectral bins using a
signal-to-noise ratio threshold, which can wash out
%absorption
weak features. We further excluded spectral regions that are subject
to calibration uncertainties in the effective area due to atomic
features.  Specifically, it is known that this calibration is poor in
the ranges $\sim$1.8--1.9 and $\sim$2.0--2.4~keV due to Si in the
detector and Au M edges in the telescope, respectively. The effective
area also has a steep change at $\sim$1.56~keV due to Al in the
telescope. Thus for the purposes of spectral fitting we conservatively
chose to exclude the 1.5--2.3~keV energy range. We also excluded the
region below 1~keV, as this suggested the need for an extra local
continuum component and does not affect our higher energy results.
%CHECK WORDING
For HXD/PIN, we first performed background subtraction
on the original 256-bin spectrum to identify the maximum
continuous spectral range with nonnegative background subtracted
counts, since
negative background-subtracted counts
would indicate an
obvious breakdown of the background model.
This was then rebinned uniformly to bin widths
of 1.5~keV, leading to the final useful spectral ranges 
12.0--35.0 and 18.0--39.0~keV for observations 702 and 706, respectively.

The XIS/PIN data relative cross-normalization involves many factors
\citep[see][for a detailed discussion]{yaqoob2012}.
Observation 702 has ``HXD-nominal'', while 706
has ``XIS-nominal'' pointing. The recommended
PIN:XIS ratios (hereafter \cpx) are then  
1.18 and 1.16, respectively\footnote{\href{ftp://legacy.gsfc.nasa.gov/suzaku/doc/xrt/suzakumemo-2008-06.pdf}{ftp://legacy.gsfc.nasa.gov/suzaku/doc/xrt/suzakumemo-2008-06.pdf}}. These
values do not take into account background-subtraction systematics,
sensitivity to spectral shape, and other factors that could affect the
actual ratio. Allowing \cpx\ to be a free parameter does not optimally
address this issue, as this could skew the best-fitting model
parameters at the expense of obtaining a \cpx\ ``best-fit'' value,
which in actuality is unrelated to the true normalization ratio of the
instruments. We thus carried out preliminary investigations for each dataset
(\scr{sec-modeling}) before deciding if we could fix \cpx. For comparison, we also performed XIS-only spectral fits.

\subsection{\nustar}
We studied four archival observations carried out by the \nustar\ mission
\citep{harrison2013}. 
%For brevity, we refer to the \nustar\ ObsIDs as 2, 4, 6, and 8.
We reduced the observations to obtain calibrated and screened level 2
event lists from the level 1 data by means of the standard
\nustar\ pipeline {\sc nupipeline}\footnote{\href{http://heasarc.gsfc.nasa.gov/docs/nustar/analysis/nustar\_swguide.pdf}{http://heasarc.gsfc.nasa.gov/docs/nustar/analysis/nustar\_swguide.pdf}, Perri et al. 2014.},
which is part of the {\sc heasoft 6.24} software package.

\vspace{.2cm}
We used the {\sc nuproducts} pipeline to extract source and background
spectra and corresponding responses for each of the two detectors
Focal Plane Module A and B (FPMA, FPMB). We chose circular source
regions centered on the source with 60\arcsec\ radii. For background
we chose rectangular regions covering essentially the full area of the
CCD, except for the source region. Initial examination of the
individual spectra showed that there were too few counts to allow
reliable fitting over a substantial spectral range. We thus
%used our
%own modified version of the {\sc ftools} task {\sc addspec} to combine
combined
all spectra and produced a master source and background spectrum for
each detector. We checked that there was minimal variability in
normalization or spectral shape between observations, so that this
process did not wash out any significant spectral features that might
be unique to a single observation.

\vspace{.3cm}
%group138-1.txt
We applied binning factors of 3 to the master source spectrum between
23.6--30~keV, and 8 above 30~keV. This was the minimal binning
scenario that ensured a minimum of 20 counts per channel for the unscaled
background, as well as both raw and background subtracted source
counts up to 43~keV both for FPMA and FPMB. In the same energy range
this choice led to a scaled background as a fraction of the background
subtracted source counts that was up to $\lesssim 50\%$ per
channel. Since \nustar\ data are not well calibrated below 3~keV, this
binning allowed the use of \cs\ statistics in the 3--43~keV energy
range.  We fitted the FPMA and FPMB spectra simultaneously,
allowing the cross-normalization to be free (see \scr{sec-nufit}).

\section{Spectral Modeling: \feka\ line emission and Reflection Continuum}\label{sec-modeling}
\subsection{Overview}
Our primary goal was to apply a physical, self-consistent model for
the \feka\ line and its associated reflected continuum that does not assume the
presence of a relativistically broadened line, and thus
determine whether the \fourc\ spectral data
can be modeled without strong gravity relativistic effects that arise
close to the central black hole at gravitational radii $R_g \lesssim
100$.  There are widespread claims in the literature that the
\feka\ line emission in these data is broadened due to such effects,
thus allowing measurements of black hole spin. In addition, Fe
abundances in previous work range from subsolar to highly supersolar
(see \scr{sec-intro}).  Instead, we tested specifically whether these
data can be fitted exclusively with narrow \feka\ line emission and
associated reflection continuum from distant matter with a finite
column density, with line and reflection continuum calculated
self-consistently, and solar Fe abundance only.  For this purpose, we
applied the toroidal \x\ reprocessor model
\myt\ \citep{murphy2009,yaqoob2012}.
 We stress that the line and
  associated reflection continuum are coupled by the atomic
  physics and are produced in tandem by the model, thus preserving self-consistency between the line and Compton-scattered continuum.
Other models that
self-consistently produce the line and reflection continuum exist but
were not used in this study. The \citet{ikeda2009} model is not
publicly available, the \citet{brightman2011} {\sc torus} has some
errors \citep[see][for details]{liu2014,balokovic2018}, while {\sc
  ctorus} \citep{liu2014} is too restrictive for our purposes.  {\sc
  borus} \citep{balokovic2018} came online after the bulk of this work
was complete. We note though that \citet{balokovic2018} reported
mostly good agreement with \myt\ (see their Appendix).

We used \xspec\ \citep[][version~12.10.0c]{arnaud1996} and the
\cs\ statistic for minimization.
We included absorption from material between the
observer and the source (\nhinter), modeled with a
{\tt phabs} component, and fixed at the
tabulated Galactic column density value of
\ten{1.16}{21}~\cunits\ \citep{kalberla2005}, unless stated
otherwise. We used photoelectric cross-sections from
\citet{verner1996} with element abundances from \citet{anders1989}.

For each parameter we calculated statistical errors for
90\%\ confidence (one interesting parameter, corresponding to
$\Delta\chi^2=2.706$), by iteratively stepping away from the best-fit
minimum. Errors for line flux and equivalent width were determined as
explained in \scr{sec-fxew}. We do not give statistical errors on
continuum fluxes and luminosities because absolute continuum fluxes
are dominated by systematic uncertainties that are not well
quantified, typically of the order of $\sim$$10\%-20\%$
\citep[e.g.][]{tsujimoto2011,madsen2017}.

%Hack to have italics small capitals in subsection title!
\subsectionM{}
\vspace{-0.55cm}
\hspace{0.39\columnwidth}{\textit{\fauxsc{mytorus} Model}}
\label{sec-mytorus}

We provide here a brief overview of salient characteristics of the
model. For in depth descriptions see
\citet{murphy2009,yaqoob2011,yaqoob2012,lamassa2014,yaqoob2016} and
the
\myt\ manual\footnote{\vspace{-.7cm}\href{http://mytorus.com/manual}{http://mytorus.com/manual}}.
The baseline geometry consists of a neutral-matter torus of circular
cross-section, with diameter characterized by the equatorial equivalent
hydrogen column
density.  A central, isotropic \x\ source illuminates the torus,
and the global covering factor of the reprocessor is 0.5,
corresponding to a half-opening angle of 60\degr. Note that the model
can also be used to mimic other geometrical configurations
\citep[][Fig.~15]{yaqoob2012} so that one is not limited to
modeling a strictly toroidal geometry. Thus in spite
of the nomenclature used for the \myt\ model and its components,
it is in fact a more general analysis tool for reprocessing
of primary \x\ continua.
The model self-consistently produces the
\feka\ and \fekb\ fluorescent emission-line spectrum, as well as
absorption and Compton scattering effects on continuum and line
emission. At present, abundances are fixed at solar values. Free
relative normalizations between different components can be used to
accommodate a variety of different actual geometries compared to the
specific assumptions in the original calculations, as well as time
delays between the various model continua and line photons. This,
however, does not break the self-consistency between fluorescent line
emission and reflection continuum.

The model's direct, line-of-sight (``zeroth'', Z) observed continuum
component is obtained from the intrinsic continuum via a
multiplicative table model ({\tt mytorus\_Ezero\_v00.fits} in \xspec)
and is not affected by the global geometry. If the angle between the
torus symmetry axis and the observer's line-of-sight (\thobs) is
greater than the torus opening angle, this continuum is diminished via
absorption and removal of photons from the line-of-sight by Compton
scattering. Further, the global distribution of matter gives rise to a
Compton-scattered (``reflected'', S) continuum and fluorescent line
(L)
emission.
The reflected continuum is implemented as an additive
table model ({\tt mytorus\_scatteredH500\_v00.fits}); this corresponds
to a power-law incident continuum with termination energy 500~keV and
photon index $1.4< \Gamma <2.6$. The \feka\ and \fekb\ emission lines
are implemented with another additive table model ({\tt
  mytl\_V000010nEp000H500\_v00.fits}). Each table has separate
parameters for incident power-law continuum normalization, photon
index $\Gamma$, angle \thobs, redshift $z$, and equivalent
hydrogen column density\footnote{For brevity, we later refer
  to the Z, S, and L tables as {\tt etable\_mytorusZ, atable\_mytorusS, atable\_mytorusL}.}. 
In general, corresponding parameters among tables are tied to each other, unless stated otherwise.
In particular, the hydrogen equivalent column density associated
  with the reflection continuum is always identical, and tied to, the
  one for the fluorescent emission line component since both this
  continuum and these emission lines are the result of Compton
  scattering due to the global matter distribution. Thus, we denote this
column density with the single symbol, \nhs.

As \fourc\ is a Type 1 AGN, we assume a face-on geometry
(\thobs~$=0\degr$) and decouple \nhs\ from \nhz, so that
\nhs\ represents the global, out of the line-of-sight column density,
while \nhz\ models the line-of-sight column density which may
have a different value. The reflection spectrum and fluorescent line
emission for inclination angles that do not intercept the torus are
similar to those for the face-on case and the differences are too
small for the data to be sensitive to them.
An illustrative sketch of the assumed configuration is shown in
  \fr{fig-NHplot}. \tr{tab-nh} summarizes the main continua, associated
column densities, terminology, and symbols used.
For a more
detailed discussion see \citet{yaqoob2012}, Section 4.5.2.

We also include a parameter $A_S$ for the relative normalization
between the direct and scattered continuum, which is 1.0 for the
baseline geometry, implying either a constant intrinsic \x\ continuum
flux or a variable one for which the \x\ reprocessor is compact enough
that the Compton-scattered flux responds to the intrinsic continuum on
time scales much less than the integration time. Values $A_s \ne 1.0$
imply departures of the covering factor from 0.5, time delays between
intrinsic and scattered continua, or both. However, the relationship
of $A_S$ to the covering factor is not simple because the detailed
shape of the scattered continuum varies with the covering
factor. Similarly, $A_L$ is the relative normalization of the
\feka\ line emission, with $A_L=1$ having a similar meaning to
$A_S=1$; we set $A_L=A_S$ throughout as otherwise the model's
self-consistency is broken.
Both parameters are implemented by
\xspec\ {\tt constant} components that multiply the S and L tables.

\vspace{-0.1cm}
\subsection{\feka\ Line Energy}
In \myt, the \feka\ line is modeled as a ${\rm K}\alpha_1$, ${\rm
  K}\alpha_2$ doublet at 6.404 and 6.391~keV with a branching ratio of
2:1, giving a weighted mean centroid energy of \erest~$=6.400$~keV.  The
\fekb\ line is centered at 7.058~keV. In \suzaku\ or \nustar\ data the
line peaks are likely to be offset due to instrumental calibration
systematics and/or mild ionization effects. We thus used the best
fit model redshift
parameter
%$z_{\rm fit}$
to calculate an effective \feka\ line energy offset
%\eshift~$=-E_{\rm rest} \frac{\Delta z}{1+\Delta z}$ where
%$\Delta z \equiv = z_{\rm fit} - z_{\rm ref}$ and $z_{\rm ref}=0.104$.
in the observed frame, so that positive shifts imply \feka\ centroid
energies higher than \erest.
%See \citet{tzanavaris2018}, Sec. 4.4 for further details.

\vspace{-1.5pt}
\subsection{\feka\ Line Velocity Width}\label{sec-gsmooth}
We implemented line velocity broadening via a Gaussian convolution
kernel ({\tt gsmooth} in \xspec) with energy width $\sigma_E =
\sigma_L \left( \frac{E_0}{6 {\rm keV}} \right)^{\alpha}$, where
$E_0$ is the centroid energy, and
\sigl\ and $\alpha$ are free parameters; we set $\alpha=1$ for a
velocity width that is independent of energy. In velocity units one
obtains ${\rm FWHM} = 2.354 c \frac{\sigma_L}{6}$, or $117,700
\sigma_L ({\rm keV})$~\kmps.
Although the reflection continuum includes edges, we did not apply
velocity broadening on it, as this significantly slows down the fitting process and previous work has shown that the effect is minimal. Even so, we tested the effect of including broadening using \texttt{gsmooth} in \xspec\ for our data. At most this induces a fractional change in parameter values of $\lesssim$3\%, the only exception being $\bm{N_{\bf H,S}}$ for obsID 706. In this case the fractional change is $\sim$20\%, however this is more than 3 to 9 times smaller than the statistical fractional error of 65\%--186\%\ (\scr{sec-nh-706}).

\vspace{-0.1cm}
\subsection{\feka\ Line Flux and Equivalent Width}\label{sec-fxew}
After the best-fit was obtained, we isolated the emission-line table
{\tt atable\_mytorusL} to measure the observed flux of the
\feka\ line, \ifeka, in an energy range excluding the \fekb\ line with
the \xspec\ {\tt flux} command. We measured the equivalent width (EW)
by means of the line flux and the total monochromatic continuum flux
at the observed line peak energy. \ifeka\ and EW in the AGN frame were
then obtained by multiplying observed values by $(1+z)$. As these are not
explicit model parameters, we estimated fractional errors by using the
fractional errors on $A_S$.
%See \citet{tzanavaris2018}, Sec. 4.6 for further details.

\vspace{-.1cm}
\subsection{Continuum Fluxes and Luminosities}
We calculated continuum fluxes and luminosities using the best-fit
model and the {\tt flux} and {\tt lumin} commands in \xspec. We
obtained absorbed fluxes in the observed frame (labeled ``obs'') and
both absorbed and unabsorbed luminosities in the AGN frame (``rest,
abso'', ``rest, unabso'', respectively). In the latter case a redshift value
of 0.104 was input to {\tt lumin}. For absorbed values, we used the total
best-fit model minus any additional gaussian emission lines, and for
unabsorbed ones only the direct power-law component. The energy
ranges used were 2--10 and 10--30~keV.

%\vspace{-.4cm}
\subsection{\suzaku\ XIS}\label{subsec-XIS}
As stated, we fitted the two observational datasets 702 and 706
independently. For each observation, we first fitted only the combined
XIS data with a baseline \myt\ model. We discuss additional components
introduced below.

Our strategy was to first step independently through each parameter
until a preliminary stable solution was found. To avoid fitting
instabilities that can be introduced by a weak emission line, we
stepped through the line redshift parameter, $z$, with \sigl\ fixed at
\ten{8.5}{-4}~keV (100~\kmps, FWHM). After a
stable minimum was obtained, $z$ was fixed and \sigl\ was left free
and stepped through to estimate the best-fit solution. If the lower
90\%\ limit tended to \sigl~$=0$, \sigl\ was frozen at 100~\kmps\ as
the narrow line was not resolved.

%This actually never happened so remove:
%If the upper 90\%\ limit was $\ga
%7000$~\kmps\ (i.e.~of the order of the CCD resolution), \sigl\ was
%fixed at the best-fit value suggested by the stepping procedure.
%

For observation 702, \nhz\ converged to a value below the lower
limit of \mytz\ ($10^{22}$~\cunits). In this column density
regime \nhz\ has negligible Compton scattering so that
simple absorption is sufficient. We thus replaced \mytz\
with {\tt zphabs} in the \myt\ model.

The only additional component to the baseline
\myt\ model that led to a statistically significant improvement
(probability that data consistent with model without extra component
less than 10\up{-4}) was \fetwofive\ emission at $\sim$6.67~keV (rest frame).
This was modeled by
a {\tt zgauss} component.
The line is unresolved and thus fixed at 100~\kmps. Such narrow
  emission lines
  are thought to originate in highly ionized gas far from the
  central SMBH and their presence is
well documented in many Type 1 AGNs and is not
surprising
\citep[e.g.][]{bianchi2004,nandra2007,bianchi2009,patrick2011,patrick2012}.

For observation 702, the full \xspec\ model was thus

\noindent{\tt phabs *\\
  (zphabs*zpowerlw\\
  +constant*atable\_mytorusS\\
  +constant*gsmooth*atable\_mytorusL\\
+zgauss)} .

For observation 706, we found that fixing the overall absorption
between the observer and the source to the tabulated Galactic value
still left a signature of excess extinction.  We thus let the
corresponding {\tt phabs} parameter free, leading to a factor of about
two increase. In addition, the $\sim$5--7~keV spectrum shows signs of
significant line-of-sight absorption larger than at lower energies. To
account for this
%we included additional
the absorption was modeled to only partially
  cover the line-of-sight. This was implemented by means of a
\texttt{zpowerlw} term for nonabsorbed emission and a
\texttt{constant*etable\_mytorusZ*zpowerlw} term for absorbed emission, where the first two components were allowed to be free, while the power law was tied to the other (nonabsorbing) power law of the model. 
Here, the {\tt constant} component is the relative contribution of
this additional power law ($A_Z$ in \tr{tab-specall}). The full
\xspec\ model was thus

\noindent{\tt phabs *\\
%  (zphabs*zpowerlw\\
  (zpowerlw\\
  +constant*etable\_mytorusZ*zpowerlw\\
  +constant*atable\_mytorusS\\
  +constant*gsmooth*atable\_mytorusL)} .

%\vspace{-.2cm}
\subsection{\suzaku\ XIS$+$PIN}
For the combined XIS$+$PIN fit, we investigated whether using the
recommended values \cpx~$=1.18$ and 1.16 (for \suzaku\ observations
702 and 706, respectively) was reasonable. We obtained
a preliminary best-fit with \cpx\ free, and then explored
\nhs~$-$~\cpx\ parameter space by means of two-dimensional contours
as shown in \fr{fig-NHS-Cpintoxis}. In this figure the horizontal dashed
lines mark the 90\%\ \nhs\ bounds from XIS-only fitting, while the
vertical dashed lines mark the fiducial recommended \cpx\ values of
1.16 and 1.18. In the case of observation 702 the upper
90\%\ \nhs\ bound intersects all three contour levels, and so does the
line for \cpx~$=1.18$. The lower \nhs\ bound intersects the
99\%\ contour. We considered this satisfactory evidence that for this
dataset the value of 1.18 was adequate and fixed the parameter
to this value for
the combined XIS$+$PIN fitting of the 702 observation. On the other
hand, for observation 706 the vertical line for
\cpx~$=1.16$ does not intersect any of the three contours, although the contours
overlap with the \nhs\ range from the XIS-only fit (which actually
only has a lower limit in \nhs). We thus left
\cpx\ free in the combined XIS$+$PIN fit.  Apart from the
cross-normalization parameter, implemented with an extra {\tt constant}
component in \xspec, the rest of the model components were as
in the XIS case for each observation.
%\todon{This still not final!}

\vspace{0.3cm}
\subsection{\nustar}\label{sec-nufit}
We fitted the FPMA and FPMB data simultaneously, allowing their
cross-normalization to be free by means of a cross-normalization
parameter \cba, implemented as a {\tt constant} component in \xspec.
We fixed this parameter to unity for FPMA, and allowed it to be free
for FPMB, thus obtaining the best-fit relative cross-normalization of
FPMB with respect to FPMA.

Our fitting strategy for obtaining a preliminary stable solution
involving the $z$ and \sigl\ parameters was as described in
\scr{subsec-XIS} for \suzaku.  As for \suzaku\ observation 702, we once more
found it necessary to replace \mytz\ with {\tt zphabs} because with
the former the \nhz\ value hit the model's lower bound.
The full \xspec\ model was thus

\noindent{\tt constant *\\
  phabs *\\
  (zphabs*zpowerlw\\
  +constant*atable\_mytorusS\\
  +constant*gsmooth*atable\_mytorusL)} .

\vspace{0.3cm}
\section{Results and Discussion}\label{sec-results}
Spectral fitting results for the \fourc\ \suzaku\ (individual
observations) and \nustar\ (combined) data with the \myt\ model are
shown in \tr{tab-specall} and Figures~\ref{fig-NuSTAR} to
\ref{fig-706-xispin}.  Column 2 in \tr{tab-specall} lists the particular best-fitting parameters or information items that are shown in subsequent columns in the same row. These are explained in detail in the table caption.  In the
Figures, panel (a) shows data and total best-fit model, (b) shows the
total and individual continuum model components, (c) the data/model
ratio, and (d) the data with total model in the vicinity of the
\feka\ line. In \tr{tab-specall} the energy shift of the \feka\ model
relative to the reference energy ($6.400 (1+z)$, $z\equiv0.104$) is in
the observed frame. The same applies to all continuum fluxes, to
facilitate comparisons with values in the literature. On the other
hand, continuum luminosities are in the AGN
(i.e.~rest) frame. All other parameters are in the AGN frame.  We
first present key points related to the spectral fitting for different
datasets. We then discuss in detail results for individual parameters.

\vspace{0.2cm}
\subsection{\nustar\ Fit}\label{sec-nustarfit}
The \nustar\ fit is shown in \fr{fig-NuSTAR}. Although we fitted the
combined FPMA and FPMB data simultaneously without merging them, for
plotting purposes and to facilitate visualization of results the FPMA
and B data and fit are shown combined. For both the FPMA and FPMB data
%the fit was drawn to somewhat high values of \sigl\ for the
%\feka\ line. At the same time however,
the lower limit for the \feka\ emission line \sigl\ was always
zero,
%, meaning that \xspec\ was trying to fit continuum.
%For \sigl~$\sim0.5$~keV, the best-fit null hypothesis probability that the data are drawn from the model was
%\pnull~\simn\ten{2}{-3}.  However this fit showed some obvious residuals at
%the line center, while the \nustar\ spectral resolution is $\sim
%12,000$ \kmps\ (FWHM). It is thus likely that the fit was driven to fit the
%continuum rather than any broad feature.
%On the other hand, a fit with
so we fixed \sigl\ at the equivalent of 100~\kmps, FWHM, consistent
with an unresolved narrow line.
%hypothesis probability%
%\pnull~\simn\ten{6}{-4}.
%This evidence suggests that quoting the second
%result might be the more conservative option.
%For reference, the \sigl\ upper limit (90\%) for this fit is 953.7~eV or
%112,208~\kmps.

\subsection{\suzaku\ Fits}
%Results of fitting the \suzaku\ data are presented in columns (3),
%(4), (5), and (6) of \tr{tab-specall}. 
%\vspace{-0.2cm}
\subsubsection{ObsID 702}\label{sec-702}
Fitting results for this ObsID are plotted in
Figures~\ref{fig-702-xis} and \ref{fig-702-xispin} for XIS and
XIS$+$PIN, respectively.  It can be seen from Columns (4) and (5) in
\tr{tab-specall} that the fitting results with and without the PIN
data are similar. For \nhs, although the best-fitting value is $\sim$7
times higher for the \xispin\ fit, within the errors both fits are
consistent with a moderately Compton thick global column density (see
also \fr{fig-NHS-Cpintoxis}, left panel).  The narrow line component
was unresolved and its width was fixed at 100~\kmps, FWHM. In
addition, in both cases we modeled emission from \fetwofive\ at
$\sim$6.67~keV (rest), which must be from a completely distinct,
highly ionized region \citep[see also,
  e.g.,][]{patrick2012,gofford2013,bianchi2005,yaqoob2003b}.

%\vspace{-0.3cm}
\subsubsection{ObsID 706}\label{sec-obs706}
Fitting results for this ObsID are plotted in
Figures~\ref{fig-706-xis} and \ref{fig-706-xispin}, respectively. As
in ObsID 702, the results with and without PIN are similar [see
  Columns (6) and (7) in \tr{tab-specall}, and \fr{fig-NHS-Cpintoxis},
  right panel]. The \nhs\ best-fit value for the \xispin\ fit is
$\sim$3 times higher and better constrained than the lower limit for
the XIS-only fit, however within the errors they are consistent with
each other, and with an overall Compton-thick column density out of
the line-of-sight.  As mentioned, we used two line-of-sight components
for this observation 
to model partial coverage. While the non-covered component has no
  absorbing column density associated with it, the covered
  component [$\sim$30\%, Columns (6) and (7) in
row (11), \tr{tab-specall}] has
  a column density of $\sim14\times10^{22}$~\cunits.
This is the only observation for which the %narrow
line component was resolved at $\sim$17,600~\kmps, FWHM. In addition, for
this observation we also found it necessary to keep the overall
intervening absorption between observer and source (\nhinter) free; it
converged to a value a factor of $\gtrsim$2$\times$ the Galactic value.

%\vspace{-0.3cm}
\subsection{Intrinsic Continuum}
In \tr{tab-specall} it can be seen that the photon index of the power law
intrinsic continuum, $\Gamma$, is similar across observations
and instruments, ranging from $\sim$1.9 to $\sim$2,
and does not appear to be sensitive either to
\cpx\ or spectral range (XIS-only versus XIS$+$PIN or \nustar).
A value of $\sim$1.9 is typical for intrinsic \x\ continua
of type 1 AGNs (e.g. \citealt{dadina2007}; \citealt{nandra2007}).

%\vspace{-0.3cm}
\subsection{\x\ Reprocessor Column Densities}\label{sec-nh-706}
The most common methodology for determining if an AGN is Compton-thick
estimates only the line-of-sight column density.
In models such as {\tt pexrav} or {\tt pexmon} the column density
associated with reflection is assume infinite; a {\it global}
column density is absent.
This is thus
the first time that the global, out of the line-of-sight, average
column density has been measured in this object. This is particularly
noteworthy as we find the two types of column density to be
significantly different. All line-of-sight column densities for all
observations are well in the Compton-thin regime, ranging from
$\sim$\ten{0.19}{22} to $\sim$\ten{14}{22} \cunits\ [row (9) in
  \tr{tab-specall}].  In contrast, all global column densities from
broadband data (\suzaku\ \xispin\ and \nustar) are unambiguously
Compton thick, ranging from $\sim$1.5 to $\sim$\ten{2.9}{24} \cunits,
while within the uncertainties even XIS-only derived ones are
consistent with being Compton thick [row (11) in \tr{tab-specall}].
%Clarify key difference from pexrav:
It is important to note also that the shape of the reflection spectrum for the global column densities found here is significantly different from that due to extremely Compton-thick matter with column densities $\gg 10^{25}$\cunits\ (which is the case for the infinite column density disk-reflection models such as {\tt pexrav} or {\tt pexmon}). It is then to be expected that our results
may differ markedly from such entirely different analyses. Only some
works that adopt such different analysis strategies have reported (single, line-of-sight) equivalent hydrogen column density values. These range from $<$$3.1\times 10^{20}$~\cunits\
\citep{fukazawa2011} to $>$$6.2\times 10^{21}$~\cunits\ \citep{gofford2013}, and would clearly lead to Compton-thin classifications.
%5x10^20 tombesi2014
%
%No neutral absorption beyond Galactic:
%larsson2008
%patrick2012

%On the other hand, the column densities of the reflected component out
%of the line-of-sight are consistent for broadband data (\nustar\ and
%\suzaku\ \xispin), giving a range of $\sim1.5$ to \ten{\sim2.9}{24} \cunits,
%in the Compton thick regime ($N_H>$~\ten{1.25}{24} \cunits).
%They are also relatively consistent between the narrower band 702 and
%706 XIS data, however suggesting values \nhs~$\gtrsim 0.4 \times 10^{24}$ \cunits\ in the Compton thin regime, and
%underscoring the importance of broadband data for properly estimating
%the level of Compton thickness in AGNs.

\subsection{Continuum Fluxes and Luminosities}
In \tr{tab-specall} the observed (absorbed) 2--10~keV fluxes are very
similar for ObsID 702 with XIS and \xispin\ (3.15
vs.~\ten{3.16}{-11}~\funits), and similarly for luminosities.  The
\nustar\ flux value is somewhat lower (\ten{2.87}{-11}~\funits), while
the ObsID 706 values somewhat lower still (\ten{2.53}{-11}
\funits). These ranges anticorrelate with the ranges in line-of-sight
column density \nhz, which is highest for ObsID 706, and lowest for
ObsID 702. The unabsorbed 2--10~keV luminosities are more similar
between \nustar\ and ObsID 702 XIS and \xispin, however ObsID 706 (XIS
and \xispin) is still lower, as in the absorbed case.  This then might
suggest that in this range the emitted flux and luminosity was
intrinsically lower for the 706 observation. This is consistent with
the findings of \citet{tombesi2014} but in disagreement with
\citet[][their Fig. 1]{lohfink2017}.

In the 10--30~keV energy range there are no similar trends either for
absorbed or for unabsorbed fluxes or luminosities. Note that the
intrinsic (unabsorbed) 10--30~keV luminosities are somewhat lower
than their absorbed counterparts, since the latter also include
reflection from the torus which increases the apparent luminosity.

\subsection{\feka\ emission line}
The best-fit energy shift of the \feka\ line peak in \tr{tab-specall}
ranges from \aer{-0.5}{+23.9}{-25.2} to \aer{16.2}{+32.9}{-35.1} eV.
The largest values are for \suzaku\ 706 XIS and \xispin\ data, and are
likely due to systematic effects in the XIS energy scale, and perhaps
some mild ionization, although all values are consistent with zero shift
within their 90\%\ errors. In any case, any such level of ionization
is negligible and the assumption of an essentially neutral reprocessor
is robust.

Looking at the zoomed-in \feka\ regions for the \suzaku\ data,
which have the highest energy resolution, in
Figures~\ref{fig-702-xis}(d) and \ref{fig-702-xispin}(d), it is
remarkable that the narrow-line \myt\ model for ObsID 702 data
provides an excellent fit to the \feka\ line, even though the line is not
resolved (see also the residuals plots (c) in the same figures).  For
the ObsID 706 data, the only dataset in which the line is resolved, an
excellent fit is also obtained with only a narrow,
although broader,
line
[Figures~\ref{fig-706-xis}(c)-(d) and \ref{fig-706-xispin}(c)-(d), and
  \tr{tab-specall}]. The best-fit \sigl\ is $\sim$150~eV or
FWHM~$\sim$17,650~\kmps.

We can obtain a rough estimate of the size of the emitting region if
we assume Keplerian motion. As in \citet{shu2011}, we assume the
line-emitting matter is virialized and orbits the black hole. The
distance $r$ to a black hole of mass \mbh\ is then given by
$GM_{\bullet}=r\langle v^2\rangle$, where $G$ is the gravitational
constant and $\langle v^2\rangle$ the velocity dispersion.  We further
assume that $v$ is related to FWHM velocity by $\langle
v^2\rangle=\frac{3}{4} \ {\rm FWHM_{Fe K\alpha}^2}$
\citep{netzer1990}, obtaining $r=\frac{4c^2}{3 \ \rm FWHM_{Fe
    K\alpha}^2} \ r_g$, where the gravitational radius $r_g\equiv
GM_{\bullet} /c^2$.  We set \mbh~$=4\times 10^9$~\msun~\citep{woo2002}
and use the \fwfeka\ range of values in \tr{tab-specall}. We only
carry out this exercise for \suzaku\ observation 706 [columns (6) and
  (7) in row 19 of \tr{tab-specall}] in which the line is resolved,
taking the upper and lower \fwfeka\ limits into account. The resulting
size range is $\sim$230 to $\sim$680~$r_g$ ($\sim$0.045 to
$\sim$0.13~pc) for the \feka\ emitting region. In addition, using the
\citet{peterson2004} relation for the size of the {\it optical}
broad line region
(BLR) for a $4\times 10^9$~\msun\ SMBH, the \citet{corbin1997} and
\citet{brinkmann1998} \hb\ emission line widths of 11,000~\kmps\ and
4,700~\kmps, FWHM, produce estimates for the radius $R_{\rm BLR}$ of 0.14
and 0.78~pc, respectively. These correspond to $\sim$740 and
$\sim$410~$r_g$, respectively. The \feka\ emitting region is then
likely to be at least partially coincident with, and possibly inside of,
the BLR.

While a range of values $\sim$230--680~$r_g$ are lower
than the {\it median} estimate of \citet{shu2011} ($3\times10^4 r_g$) for
narrow \feka\ AGNs, it is well beyond the strong gravity regime,
which would require less than $\sim$50~$r_g$
\citep[e.g.][]{fabian1989,reynolds2003}. In addition, the estimates
for the relative sizes of the \feka\ and optical broad line emitting regions
are entirely consistent with \citet{shu2011} who found that
the narrow \feka\ line is likely to arise in a region within
a factor of $\sim$0.7--11$\times R_{\rm BLR}$.
The fact that, compared to the rest of the observations, the line
appears broader in this one may be a combination of several effects.
It is well known in the optical that AGN emission lines can change
even as far as leading to Type 1 AGNs being reclassified as Type 2s
\citep[``changing look'' AGNs, e.g.][and references
  therein]{lamassa2015}.  In the X-rays, the \feka\ line shape, which
depends on the global distribution of scattering material,
cannot change only under the effect of motion, since, overall, the same amount
of matter should be present between observations. Rather, the
observed change should be due to thermodynamics and/or ionization effects,
which control the amount of neutral material. A change in the
average distance of fluorescent and
scattering matter will have an effect in this
respect, as with varying distance the level of ionization will
change. Further, the broadest part of the narrow line may come from
the outer parts of the disk, perhaps a wind for which we know that
both amount of matter and ionization can be variable. Note also that
this observation combines the longest \suzaku\ exposure time and
highest spectral resolution, while fixing the width in the other
observations, where the line is not resolved, to 100 km/s is somewhat
arbitrary. The line could be broader than that (but still below the
resolution limit) so that, in fact, the discrepancy with the 706 width
might be smaller.

Finally, the line fluxes and EWs are all consistent with each other
across instruments and observations, ranging from
\aer{1.76}{+0.21}{-0.20} to \aer{1.99}{+0.87}{-0.44}
$\times10$\up{-5}~\phunits\ and \aer{38}{+5}{-4} to \aer{51}{+40}{-12} eV,
respectively.

For observation 706, the resolved \feka\ line is somewhat broad;
we also estimated above that the emission may at least 
\clearpage
\pagebreak[4]
\global\pdfpageattr\expandafter{\the\pdfpageattr/Rotate 90}
\renewcommand{\arraystretch}{1.1}
\input{tab-specall.tex}
\renewcommand{\arraystretch}{1}
{\pagebreak[4]\global\pdfpageattr\expandafter{\the\pdfpageattr/Rotate 0}}

\noindent partially be
occurring inside the BLR.

 The emission might then be coming from a
  truncated accretion disk, as suggested by \citet{tombesi2014}, who
  modeled this observation with a relativistally blurred ({\tt
    kdblur}) \pexmon\ component. \citet{bhatta2018} also reported
  evidence for a truncated accretion disk after modeling the longest
  \nustar\ observation with \relxill. To test this hypothesis, we
  modeled the \xispin\ data of this observation with
  \relxill\ \citep[][version~1.2.0]{garcia2014,dauser2014} and the
  model setup of \citet{bhatta2018},
  i.e.~\texttt{constant*tbabs*warmabs*relxill}, where as before the
  \texttt{constant} component accounts for XIS vs.~PIN
  cross-normalization. We adopted the \citet{bhatta2018} assumptions
  for Galactic column density (\ten{2.31}{21}\cunits) and
  \texttt{warmabs}, i.e.~outflow velocity of 3,600~\kmps, column
  density \ten{3.5}{21}~\cunits, ionization parameter 2.6 in the log,
  and turbulent velocity 100~\kmps. For the purposes of comparison
  with these two previous works we assumed a uniform emissivity index
  of 3, a maximally spinning black hole, solar iron abundance, and cutoff
  energy of 180~keV.
  %
  %relxill_Tombesi.xcm
  We first imposed the further assumptions of an outer disk radius
  $R_{\rm out}=400 r_g$ and disk inclination angle
  30\degr\ \citep{tombesi2014}.  This produced a best fit
  with an inner disk radius of
  $R_{\rm in}=$~\aer{32}{+14}{-12}~$r_g$ and reflection fraction
  $R=$~\aer{0.41}{+0.05}{-0.04}, consistent with the
  \citet{tombesi2014} results. However, the fit is rather poor
  (\csn~$=1.3$, \pnull~$=$~\ten{3}{-4}).
  %
  %relxill_bhatta_4.xcm
  Modifying the model setup by
  fixing $R_{\rm out}=1000\, r_g$ and disk inclination angle
  to 45\degr\ \citep{bhatta2018} only produced a poor best fit, with
  \csn~$=1.5$ and \pnull~$=$~\ten{1}{-8}.
  %
  %relxill_freeq_Bhatta.xcm
  Modifying this model setup
  by relaxing the requirement for uniform emissivity leads to an
  improved fit (\csn~$=1.2$, \pnull~$=0.01$) with
  an inner disk emissivity index of \aer{5}{+2}{-1} and an outer disk
  index of \aer{0.8}{+0.5}{-1.1} for an assumed break radius of
  $20 \, r_g$. This result is however not consistent with a truncated disk
  ($R_{\rm in}=$~\aer{2.2}{+0.6}{-0.3}~$R_{\rm ISCO}$).
  We note that this result is at least qualitatively
  consistent with theoretical modeling predictions that favor a much
  steeper emissivity profile in the inner accretion disk
  \citep{wilkins2011,wilkins2012,gonzalez2017}.

In conclusion,
relatively simple, standard \myt\ configurations that model
a narrow \feka\ emission line at 6.4~keV and its associated
reflection continuum originating
in neutral matter far from the accretion disk with solar
Fe abundance are entirely
able to provide excellent fits to all the data.
None of the observations provide any
statistical evidence for an
additional
relativistically broadened \feka\ emission
line.
The \suzaku\ 706 observation, where a $\sim$17,000~\kmps\ emission line is resolved, can however alternatively be modeled
  exclusively by a relativistically broadened component, consistent
  with a maximal black hole spin value, as in earlier results.

To our knowledge, this is the first time that data for \fourc\ were
fitted with narrow-only \feka\ emission, using a physically motivated
model such as \myt, which self-consistently models both \feka\ line
emission and the reflection continuum, and uses only solar abundances.
In addition, previous works that have analyzed these \suzaku\ and
\nustar\ data commonly start with the assumption that there is a broad
component of the \feka\ line, motivated by simple power law residual
plots (e.g.~Fig.~5 in \citealt{larsson2008} and Fig.~3 in
\citealt{patrick2012} for \suzaku\ ObsID 702; Fig.~2 in
\citealt{lohfink2017} for the average FPMA/B \nustar\ data; see also
Fig.~1 in \citealt{ballantyne2005} for \xmm\ data). They then proceed
to model the broad line by including relativistic components, which can
lead to estimates of black hole spin \citep{lohfink2017}.
Any additional narrow \feka\ component
is routinely added as an additional ad-hoc gaussian, which is completely different to the self-consistent \feka\ line emission and associated reflection
continuum of \myt. Our starting point was entirely
different in that we first 
modeled the simplest, less exotic case of
narrow line \feka\ emission to assess how well it can fit the data without
a broad component. If there were a prominent broad feature, such a model should
fail to provide even a moderately good fit. We found that this is not the
case, contradicting previous results.

%Following the \citet{yaqoob2016} results for
%Fairall~9, this is then the second case of a well-known AGN,
%previously claimed to be fitted with relativistically broadened
%\feka\ line emission, leadin to black hole spin measurements, that is
%now modeled exclusively with narrow \feka\ line emission, an
%associated reflection continuum, and solar abundances.

The particular case of \suzaku\ ObsID 702 is noteworthy. The majority
of previous analyses of this ObsID found a broad line, with $\sigma$
values reported as high as $240-300$, $<$190, and $\sim$270 eV
\citep[][respectively]{larsson2008,patrick2012,gofford2013}. This
would mean that \suzaku\ resolved the line.  In contrast, our analysis
shows that the line is not resolved, and our fitting process
(\scr{subsec-XIS}) justified fixing the line to a mere 100~\kmps, FWHM,
leading to an excellent fit as is evident in
Figures~\ref{fig-702-xis}(d) and \ref{fig-702-xispin}(d).

Given that the \suzaku\ data, which have higher energy resolution than
\nustar, do not support a broad-line component, and that the lower
\sigl\ limit for the \nustar\ data is zero, our results are unable to
confirm those of \citet{lohfink2017} who report \sigl\ of
\aer{0.57}{+0.09}{-0.08}~keV for the same \nustar\ data. We are then
forced to conclude that such a value might be an artefact of data
quality that leads to a broad-line model component partly modeling the
continuum instead of a bone-fidae broad line.

\section{Summary and Conclusions}\label{sec-summ}
We extracted \x\ spectra from two \suzaku\ and four merged
\nustar\ \x\ observations of the quasar \fourc\ to test
the hypothesis that these \x\ spectra can be fitted exclusively
with a narrow \feka\ emission line and associated reflected continuum.
A prominent broad line has been claimed in several earlier works, including
some that are based on the same data as in the present work,
and this Type 1 AGN has a relatively simple
\x\ spectrum, making it an ideal test case.

We fitted the spectra
with \myt\ that self-consistently models the \feka\ line emission and
associated reflected continuum from finite column-density material
with solar Fe abundance. We fitted \nustar\ FPMA and FPMB data
simultaneously with free cross-normalization, and estimated the best
cross-normalization factor \cpx\ for \suzaku\ \xispin\ data. Our key results
are as follows:
\vspace{-.3cm}
\begin{enumerate}
\itemsep-4pt 
\item For all data, regardless of telescope or instrument, we obtained
  excellent fits (\csn~$\sim$1) with narrow-only \feka\ line emission and associated reflected continuum.
\item Our fits require only solar Fe abundance in an X-ray reprocessor
  with finite column density, far from the central SMBH.
\item For the first time, we measured the global column density
  associated with Compton scattering out of the line-of-sight (\nhs)
  independently of the line-of-sight column density (\nhz).  For all
  observations, while \nhs\ is in the Compton thick regime
  ($\sim$1.5 to $\sim$$2.9\times10^{24}$\cunits), \nhz\ is Compton
  thin ($\sim$\ten{0.19}{22} to $\sim$\ten{14}{22} \cunits). This has
  important implications for estimates of the fraction of
  Compton-thick AGNs in the universe, which routinely use only
  line-of-sight equivalent hydrogen column density values.
\item The \feka\ line is not resolved, with the exception of \suzaku\ ObsID 706 (FWHM $\sim$17,600~\kmps).
This observation can alternatively be modeled with
a relativistic model with a broken power law emissivity. However,
two different model setups for a truncated accretion disk are not favored by the data.
\item Simple estimates suggest the size of the \feka\ emitting region
  is $\sim$230 to $\sim$680 gravitational radii from the central SMBH,
  well beyond the strong gravity regime.
\item Our results suggest that \x\ reprocessing in these data does not arise in
  the strong gravity regime in the inner part of the accretion disk, and
  thus cannot constrain black hole spin.
%\item The \suzaku\ ObsID 706 (XIS, \xispin) data require an
%  intervening column density that is about twice the tabulated
%  Galactic value, and a partial covering line-of-sight
%  component. Otherwise, the data are fitted with a relatively standard
%  \myt\ setup.
\end{enumerate}

For the \suzaku\ ObsID~706 data, further modeling of a possibly truncated accretion disk, motivated by the lower-limit estimate for the size of the \feka\ emitting region, is beyond the scope of this paper, which explores if a purely nonrelativistic model such as \myt\ is able to fit these data. We do note though that the great majority of previous claims for relativistic broadening are based on the \suzaku\ ObsID~702 and \nustar\ data (see Appendix~\ref{app-prev}). Our modeling results show that in these cases there is no hint of even a moderately broad line, thus providing no support for even exploratory relativistic modeling. 

Fairall~9, is another AGN that was widely thought to show a
relativistically broadened \feka\ emission line in its \x\ spectra
until it was recently modeled successfully only with a narrow line and its
associated reflected continuum, all with solar Fe
abundance \citep{yaqoob2016}. In the case of
\fourc\ only two previous works (\citealt{brinkmann1998} and
\citealt{noda2013}) reported narrow-only \feka\ line emission. Given
the far-reaching implications of detecting \feka\ emission originating
in the inner part of the accretion disk, the possibility of less
exotic, mundane modeling with narrow \feka\ emission should carefully
be considered, while the case for SMBH spin measurements remains
open.

\acknowledgments

\noindent We thank the anonymous referee for his/her constructive
comments that helped improve the paper. P.T. acknowledges support from
NASA grant 80NSSC18K0408 (solicitation NNH17ZDA001N-ADAP).  This work
made use of data from the \nustar\ mission, a project led by the
California Institute of Technology, managed by the Jet Propulsion
Laboratory and funded by the National Aeronautics and Space
Administration. This research made use of the \nustar\ Data Analysis
Software (NuSTARDAS) jointly developed by the ASI Science Data Center
(ASDC, Italy) and the California Institute of Technology (USA). This
research has made use of data obtained from the \suzaku\ satellite, a
collaborative mission between the space agencies of Japan (JAXA) and
the USA (NASA).
%\bm{$z\sim1$}

\noindent\facilities{\suzaku, \nustar}
\vspace{-0.5cm}
\bibliographystyle{likeapj}
\bibliography{masterbib}   

%\clearpage

\appendix
%\numberwithin{figure}{section} %needs amsmath!

\section{\\Overview of Previous Results}\label{app-prev}
For reference, we present here a further detailed overview of previous
results for \fourc.

The object features prominent $10^{\prime}$ radio lobes
\citep{riley1989} and a one-sided radio jet \citep{riley1990}.  Based
on the jet asymmetry at parsec scales, \citet{pearson1992} estimated a
jet inclination of $\approxlt$$49\degr$ to the line-of-sight.

\feka\ line emission was first detected with
\asca\ \citep{brinkmann1998,sambruna1999,reeves2000}, and later with
{\it BeppoSAX} \citep{hasenkopf2002}.
While \citet{brinkmann1998} detected only a narrow line $\sigma$$\sim$90~eV
with \asca, \citet{sambruna1999} reported a broad line with
$\sigma$$\sim$590~eV and EW$\sim$215~eV.
For all \asca\ and {\it BeppoSAX} observations,
\citet{hasenkopf2002} quote EW$\sim$200~eV with factor of $\sim$2
uncertainties and strong Compton reflection components.
             
In more recent results, \citet{ballantyne2005} presented evidence that
the \xmm\ 2004 EPIC-pn observation shows a broad (EW$\sim$130--300~eV)
ionized \feka\ line extending very close to a maximally spinning black hole.
Their best-fit model (2--12~keV)
was a relativistically blurred ionized disk with an unblurred neutral
reflector \citep{ross1999} and an additional narrow Gaussian emission
line.  \citet{ballantyne2005b} extended this analysis down to 0.3~keV,
and also reported evidence for excess cold absorption compared to the
Galactic value, as well as warm absorption, but no soft excess.

\citet{larsson2008} analyzed a \suzaku\ observation \citep[obs. ID
  702057010; ``B'' in][]{tombesi2014}, finding
evidence for $\sim$20\%\ flux increase during the observation. Their
best-fitting model for ionized reflection with relativistic blurring
reported broad (EW~$\gtrsim80$ eV) \feka\ line emission originating beyond
$50 r_g$ and a line-of-sight inclination of
$\sim$$20\degr$. This led to the suggestion of an inner truncated
accretion disk. They further concluded that a narrow component was not
required since it lead to only a marginal improvement of their fit.

For the same \suzaku\ observation, \citet{patrick2012} fitted a broad
\feka\ at 6.1~keV with a Gaussian ($\sigma<0.19$~keV, EW~$\sim22$
eV), and also found \fetwofive\ emission. However, their
{\tt relline} fit was unable to constrain black hole spin.
\citet{gofford2013} also analyzed
this \suzaku\ observation.
%Table D4
Using \reflionx, they fitted a narrow \feka\ ($\sigma=10$ eV) at 6.33
keV; they fitted broad \feka\ line emission with a Gaussian ($\sigma \sim
270$~eV). The EWs for these were $\sim$13 and 47~eV, respectively.
%Table D2
They included two warm absorbers from \xstar\ modeling, and detected a
single absorption trough at $E>8$~keV, which they interpreted as an
Ultra-Fast Outflow (UFO, $\sim0.2c$), although the specific ion
identifications may be ambiguous (``degenerate \xstar\ solutions''),
and the statistical detection is marginal.

\citet{noda2013} also analyzed this observation, reporting soft
\x\ absorption above the Galactic value.
%PT Difficult to compare with their fitting as it is done in
%steps, i.e.~no full final fit that includes everything.
They fitted time resolved soft and hard excess emission, as well as
the time-averaged spectra. Although they obtained good fits with
several phenomenological models, they concluded that relativistically
smeared ionized reflection ({\tt kdblur*reflionx+pexrav}) could not
provide an adequate model for the excesses.
%PT They say this in the conclusions, but on p. 5 they
%say that any of the 5 models, including this will
%fit all SSE spectra!
Instead, for their sample, they concluded that the soft excess was
consistently well-modeled by a thermal comptonization component
separate from the main power law continuum.  Thus the strong excess
was deemed likely to arise in a distant cold reflector.  The
\feka\ line had $\sigma=10^{-4}$~keV (fixed) and EW~$\sim$$30-37$ eV,
depending on the model.

\citet{tombesi2014} analyzed the \xmm/EPIC-pn (``A'') and another
\suzaku\ observation (``C'', 706028010).
For obs.~A, they used {\tt xillver} to
fit \feka\ line emission at $\sim$6.5~keV as well as an ionized reflection component,
and \xstar\ tables for two warm absorber components. They fitted
absorption at \erest~$\sim$7.3~keV with a Gaussian, and identified it
as due to an Ultra Fast Outflow (UFO, $\sigma_v < 12,000$~\kmps,
$v_{\rm out} = 13,500 \pm 2,400$ \kmps). They obtained several possible
best-fits for obs.~C, including (1) neutral \feka\ line emission (\pexmon)
with two warm absorbers (\xstar), and broad absorption consistent with
a UFO; (2) redshifted broad neutral \feka\ line emission from reflection
off the accretion disk (\pexmon+{\tt kdblur}) consistent with a
truncated disk; (3) ionized partial covering absorption ({\tt zxipcf}).

\citet{digesu2016} fitted the \xmm/EPIC-pn observation with a
phenomenological model, detecting excess soft \x\ absorption compared
to the Galactic value. They further performed a combined
\chandra\ High Energy Transmission Grating Spectrometer (HETGS) and
\xmm-Reflection Grating Spectrometer (RGS, 0.4--2.0~keV) spectral
analysis. The HETGS data came both from the Medium (MEG, 0.7--6.2
keV) and the High Energy (HEG, 2.5--8.3~keV) grating. For their joint
HETGS+RGS fit, absorption components were tied but continua were left
to vary. The model was a phenomenological modified blackbody with a
broad Gaussian \feka\ line emission line (FWHM$\sim$0.16\AA, consistent
with other \xmm-based results). Several soft \x\ absorption features
were also included, and a highly ionized warm absorber was also fitted,
corresponding to an outflow velocity of $\sim$3600~\kmps.

\citet{lohfink2017} performed a joint analysis of \nustar\ with
simultaneous \swift/XRT snapshots. Their best model included
\pexmon\ for cold reflection, and {\tt relxilllpCp} combining
relativistically blurred ionized reflection and a thermal
Comptonization continuum. They measured a high-energy cutoff of
\aer{183}{+51}{-35}~keV, and reported ionized reflection, a mildly
ionized warm absorber, excess cold absorption $<$2$\times
10^{21}$~\cunits, and a broad \feka\ line with equivalent width
$\sim$200~eV and $\sigma=$~\aer{0.57}{+0.09}{-0.08}~keV.
They reported spin values constrained to $>$0.5.

\citet{bhatta2018} modeled the longest ($\sim$90~ks)
  \nustar\ observation with \relxill\ for a maximally spinning SMBH,
  with a cut-off energy of 180~keV and solar iron abundance.
  They obtained an inner accretion disk radius of $\sim$35$R_{\rm
      ISCO}$, suggesting a truncated accretion disk.   

\end{document}

%% file: newcomm.tex
\newcommand\fauxsc[1]{\fauxschelper#1 \relax\relax}
\def\fauxschelper#1 #2\relax{%
  \fauxschelphelp#1\relax\relax%
  \if\relax#2\relax\else\ \fauxschelper#2\relax\fi%
}
\def\Hscale{.85}\def\Vscale{.74}\def\Cscale{1.12}
\def\fauxschelphelp#1#2\relax{%
  \ifnum`#1>``\ifnum`#1<`\{\scalebox{\Hscale}[\Vscale]{\uppercase{#1}}\else%
    \scalebox{\Cscale}[1]{#1}\fi\else\scalebox{\Cscale}[1]{#1}\fi%
  \ifx\relax#2\relax\else\fauxschelphelp#2\relax\fi}
\usepackage{textgreek} %no math mode - just \textalpha etc
\usepackage{upgreek} %in math mode - just $\upalpha$ etc

\newcommand{\fourc}{4C~74.26}
\newcommand{\mstf}{MCG~6$-$30$-$15}

\newcommand{\cunits}{cm$^{-2}$}
\newcommand{\funits}{erg~cm$^{-2}$~s$^{-1}$}

\newcommand{\kmps}{km s$^{-1}$}

\newcommand{\lunits}{erg~s$^{-1}$}

\newcommand{\mbh}{$M_{\bullet}$}

\newcommand{\msun}{$M_{\odot}$}

\newcommand{\phunits}{photons~cm$^{-2}$~s$^{-1}$}

\newcommand{\cs}{$\chi^2$}

\newcommand{\csn}{$\chi^2_\nu$}

\newcommand{\erest}{$E_{\rm rest}$}
\newcommand{\eshift}{$E_{\rm shift}$}
\newcommand{\ewfeka}{$\rm EW_{Fe K\alpha}$}

\newcommand{\fwfeka}{$\rm FWHM_{Fe K\alpha}$}

\newcommand{\lbol}{$L_{\rm bol}$}

\newcommand{\lxht}{$L_{X,{\rm 2.0-10.0 \ keV}}$}

\newcommand{\nhinter}{$N_{\rm H}^{\rm inter}$}

\newcommand{\nhz}{$N_{\rm H,Z}$}

\newcommand{\nhs}{$N_{\rm H,S}$}

\newcommand{\pnull}{$P_{\rm null}$}

\newcommand{\sigl}{$\sigma_L$}

\newcommand{\x}{X-ray}

\newcommand{\ifeka}{$I_{\rm Fe K\alpha}$}

\newcommand{\lcra}{$L_{\rm 2-10, c, rest, abso}$}
\newcommand{\lcru}{$L_{\rm 2-10, c, rest, unabso}$}
\newcommand{\lcraH}{$L_{\rm 10-30, c, rest, abso}$}
\newcommand{\lcruH}{$L_{\rm 10-30, c, rest, unabso}$}

\newcommand{\fcobs}{$f_{\rm 2-10, c, obs}$}
\newcommand{\fcobsH}{$f_{\rm 10-30, c, obs}$}

\newcommand{\thobs}{$\theta_{\rm obs}$}

\newcommand{\feka}{Fe K$\alpha$}

\newcommand{\fekb}{Fe K$\beta$}

\newcommand{\fetwofive}{Fe{\sc \,xxv}}

\newcommand{\hb}{H$\beta$}

\newcommand{\asca}{{\it ASCA}}
\newcommand{\chandra}{\textit{Chandra}}
\newcommand{\cba}{$C_{\rm FPMB:FPMA}$}
\newcommand{\cpx}{$C_{\rm PIN:XIS}$}
\newcommand{\ccross}{$C_{\rm cross}$}
\newcommand{\nustar}{\textit{NuSTAR}}
\newcommand{\suzaku}{\textit{Suzaku}}

\newcommand{\swift}{{\it Swift}}

\newcommand{\xmm}{\textit{XMM-Newton}}
\newcommand{\xispin}{XIS$+$PIN}

\newcommand{\myt}{{\textsc{mytorus}}}

\newcommand{\mytz}{{\sc my}torus{\sc z}}
\newcommand{\pexmon}{{\texttt{pexmon}}}

\newcommand{\reflionx}{{\tt reflionx}}
\newcommand{\relxill}{{\texttt{relxill}}}

\newcommand{\xspec}{{\textsc{xspec}}}
\newcommand{\xstar}{{\sc xstar}}

\newcommand{\fr}{Figure~\ref}

\newcommand{\scr}{Section~\ref}
\newcommand{\tr}{Table~\ref}
\newcommand{\exi}{\begin{equation}}
\newcommand{\exo}{\end{equation}}

\newcommand{\aer}[3]{$#1^{#2}_{#3}$} %1.2^+0.1_-0.2
\newcommand{\ten}[2]{$#1\times 10^{#2}$} %{some number} x 10^{some power}
\newcommand{\up}[1]{$^{#1}$}

\def\spose#1{\hbox to 0pt{#1\hss}} 
\def\approxlt{\mathrel{\spose{\lower 3pt\hbox{$\sim$}}
        \raise 2.0pt\hbox{$<$}}}
\def\approxgt{\mathrel{\spose{\lower 3pt\hbox{$\sim$}}
        \raise 2.0pt\hbox{$>$}}}

%% file: tab-suznu.tex
\begin{deluxetable*}{ccccc ccc c}
  \tablecaption{Exposure times and count rates for \suzaku\ and \nustar\ spectra.\label{tab-suznu}}
\tabletypesize{\small}
\tablehead{
\colhead{Telescope} &
\colhead{ObsID} &
\colhead{Date/Time} &
\colhead{Detector} &
\colhead{Exposure} &
\colhead{Energy ranges} &
\colhead{Count Rate} &
\colhead{Percentage of} &
\colhead{Refe-}
\\[-.15cm]
\colhead{} &
\colhead{} & 
\colhead{} & 
\colhead{} &
\colhead{(ks)} &
\colhead{(keV)} &
\colhead{(count s\up{-1})} &
\colhead{on-source rate} &
\colhead{rences}
\\[-.6cm]}
\colnumbers
\startdata
%dmkeypar  4C7426_702057010_xi013_src.pi exposure echo+
%274757.1  → /3 = 91.6 
%dmkeypar  4C7426_702057010_hxd_pin_sr_grp4.pi exposure echo+
%87335.75
\mrw{\suzaku} & \mrw{{\bf702}057010} & \mrw{2007-10-28T10:21:17} & XIS & 91.6  & $1.0-1.5,\ 2.3-9.5$  & $1.0230\pm0.0020$ & 98.3 & \mrw{1, 2, 3, 4, 5} \\
              &                      &                           & PIN & 87.3  & $12.0-35.0$          & $0.1115\pm0.0025$ & 22.0 &  \\                
\mrw{\suzaku} & \mrw{{\bf706}028010} & \mrw{2011-11-23T12:58:54} & XIS & 101.4 & $1.0-1.5,\ 2.3-10.0$ & $0.9601\pm0.0018$ & 98.2 & \mrw{2}\\
              &                      &                           & PIN & 109.7 & $18.0-39.0$          & $0.0528\pm0.0015$ & 23.6 & \\                
%702057010 & 2007-10-28 & XIS & ????  & $1.0-1.5, 2.3-9.5$  &  &  \\
%          &            & PIN &       & $12.0-35.0$         &  &  \\                
%706028010 & 2011-11-23 & XIS & ????  & $1.0-1.5, 2.3-10.0$ &  &  \\
%          &            & PIN &       & $18.0-39.0$         &  &  \\
\mrw{\nustar} &\mrw{6000108000{\bf2}} & \mrw{2014-09-21T15:21:07} & FPMA & \mrw{19.1}  & \mrw{$3-43$} & $0.7658\pm0.0065$ & 95.6 & \mrw{6}\\
              &                       &                           & FPMB &             &              & $0.7290\pm0.0063$ & 95.6 & \\     
\mrw{\nustar} &\mrw{6000108000{\bf4}} & \mrw{2014-09-22T11:51:07} & FPMA & \mrw{56.6}  & \mrw{$3-43$} & $0.8022\pm0.0039$ & 95.7 & \mrw{6}\\
              &                       &                           & FPMB &             &              & $0.7652\pm0.0038$ & 95.7 & \\         
\mrw{\nustar} &\mrw{6000108000{\bf6}} & \mrw{2014-10-30T23:06:07} & FPMA & \mrw{90.9}  & \mrw{$3-43$} & $0.7035\pm0.0028$ & 95.3 & \mrw{6}\\
              &                       &                           & FPMB &             &              & $0.6833\pm0.0028$ & 95.4 & \\        
\mrw{\nustar} &\mrw{6000108000{\bf8}} & \mrw{2014-12-22T06:16:07} & FPMA & \mrw{42.8}  & \mrw{$3-43$} & $0.7171\pm0.0042$ & 95.5 & \mrw{6}\\
              &                       &                           & FPMB &             &              & $0.6916\pm0.0041$ & 95.5 & \\
\mrw{\nustar} &\mrw{total}            &                           & FPMA & \mrw{209.4} & \mrw{$3-43$} & $0.7133\pm0.0018$ & 99.6 & \mrw{6}\\
              &                       &                           & FPMB &             &              & $0.6882\pm0.0018$ & 99.6 & \\
\enddata
  \renewcommand{\baselinestretch}{0.5} 
\tablecomments{Column (3) is the observation start-date (header keyword {\sc date-obs}).
  Column (5) is the exposure time (header keyword {\sc exposure}).
  Column (7) is the background-subtracted count rate
  in the energy bands specified. For the XIS, this is the rate per XIS
  unit, averaged over XIS0, XIS1, and XIS3. Column (8) is the
  background-subtracted source count rate as a percentage of the total
  on-source count rate, in the energy intervals shown. References: 1: \citet{larsson2008}; 2: \citet{tombesi2014};
  3: \citet{patrick2012}; 4: \citet{gofford2013}; 5: \citet{noda2013}; 6: \citet{lohfink2017}.
}
  \renewcommand{\baselinestretch}{1} 
\vspace{-1cm}
\end{deluxetable*}

%% file: fig-NHplot.tex
\begin{figure*}[t!]
  \vspace{-0.7cm}
\plotone{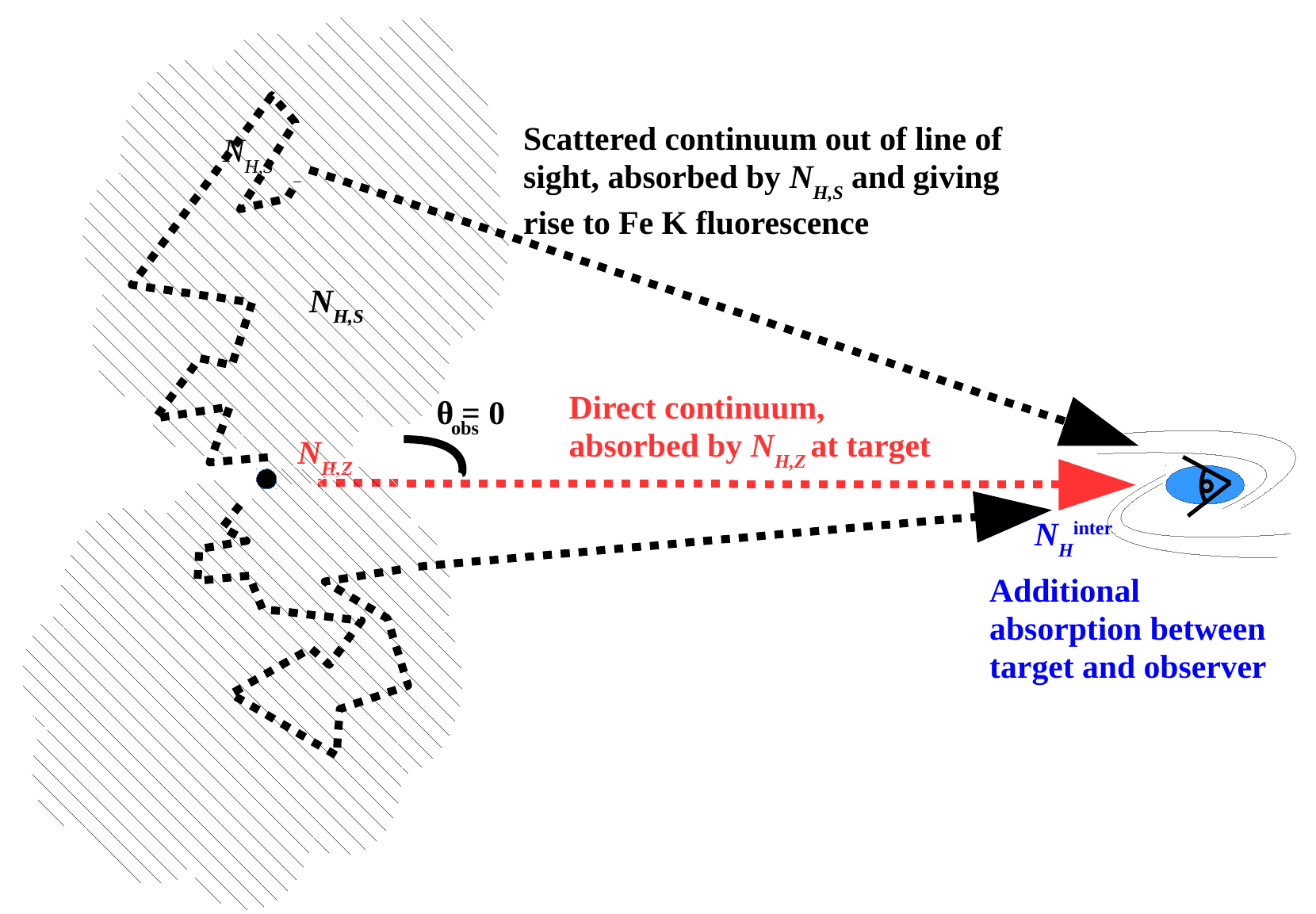}
\caption{\footnotesize
Illustration of the assumed geometry in this paper (see also \tr{tab-nh}). Part of the intrinsic continuum experiences no Compton scattering and reaches the observer along the line-of-sight, in the process being attenuated by an equivalent hydrogen column density \nhz\ (``direct'' or ``zeroth'' or line-of-sight continuum). Another part of the intrinsic continuum experiences Compton scattering out of the line-of-sight, and is absorbed by a column density \nhs, which also gives rise to the \feka\ and $\beta$ fluorescent line emission. Since \nhs\ is associated with any location out of the line-of-sight, it is a ``global'' property. A cross-section of the putative torus is drawn in the plane of the paper and is for illustration only. In reality X-ray reprocessing may occur in a collection of clouds or clumps.
    \label{fig-NHplot}
    }
\end{figure*}

%% file: fig-NHS-Cpintoxis.tex
%\begin{figure}
%  \hspace{0cm}
%  \includegraphics[width=.65\textwidth]{fig-NHS-Cpintoxis.pdf}
%  \vspace{-1cm}
%  \caption{\nhs\ from XIS+PIN \myt\ fitting against \cpx\ for
%    observation 702.  The horizontal lines mark the 99\%\ bounds from
%    the XIS-only best fit, and the vertical lines the recommended
%    \cpx\ values of 1.16 and 1.18. \todon{The best-fit is not the one
%    in the table!}}
%  \label{fig-NHS-Cpintoxis}
%  \vspace{-.2cm}
%  \end{figure}
\begin{figure*}[t!]
  \vspace{-1.4cm}
    \begin{minipage}[c]{0.5\textwidth}
        \includegraphics[width=1.2\textwidth]{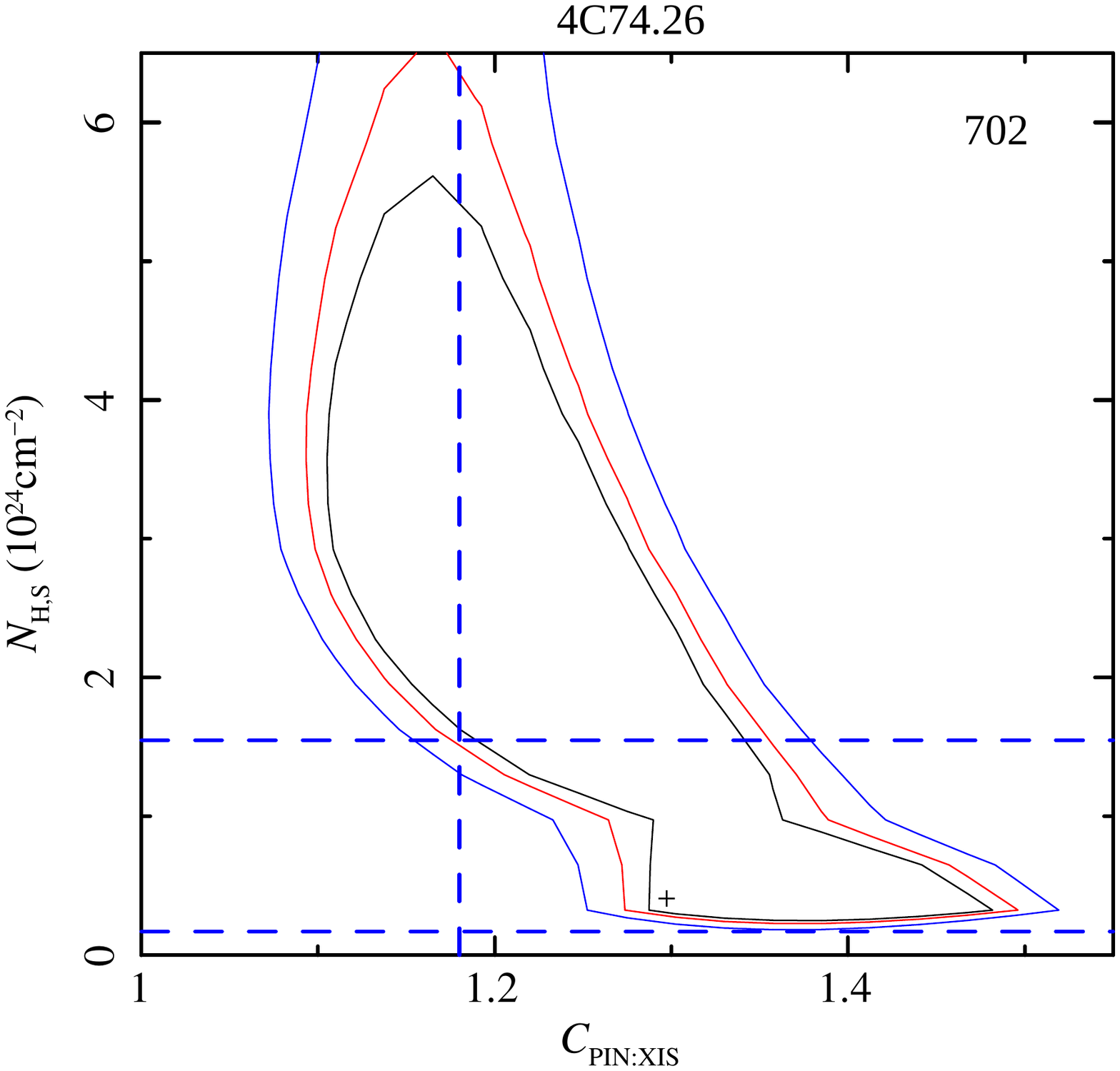}
    \end{minipage}
    \begin{minipage}[c]{0.5\textwidth}
        \includegraphics[width=1.2\textwidth]{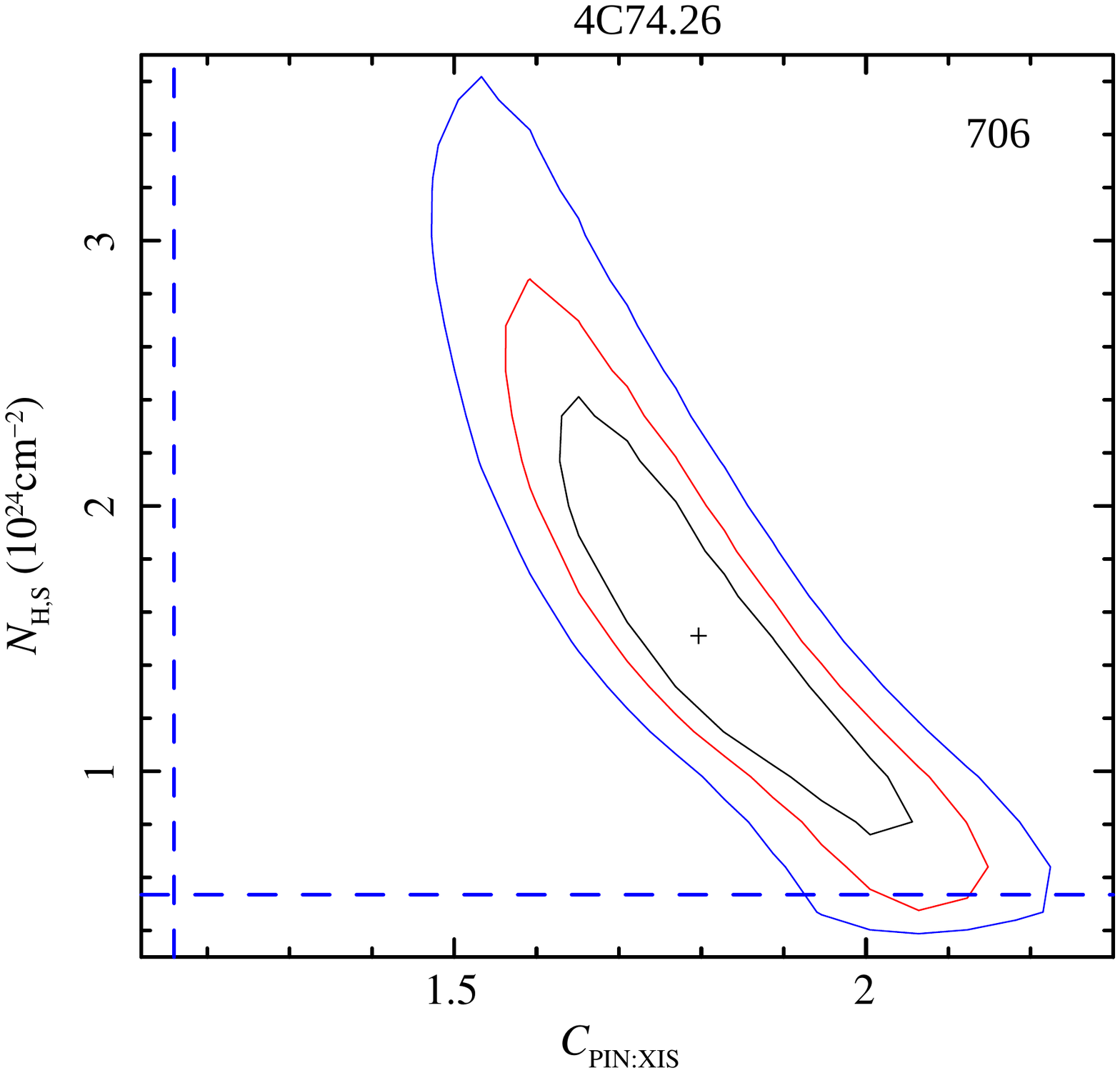}
    \end{minipage}
    \vspace{-.6cm}
  \caption{\footnotesize Determination of optimal cross-normalization, \cpx, for
    \suzaku\ \xispin\ fits. Shown are contours of \nhs\ from XIS$+$PIN
    \myt\ fitting against \cpx\ for observations 702 and 706.  The
    horizontal lines mark the 90\%\ bounds from the XIS-only best fit
    (only lower bound for 706), and the vertical lines the recommended
    \cpx\ values of 1.18 (for 702) and 1.16 (for 706). Contours are
    shown at 68\%, 90\%, and 99\%\ confidence.
    %\todon{The best-fit left is not the one in the table!}
  \label{fig-NHS-Cpintoxis}}
\end{figure*}

%% file: tab-nh.tex
\begin{deluxetable*}{lccl}
  \vspace{-0.5cm}
  \tablecaption{\bf Summary of continua, associated equivalent hydrogen column densities, and related terminologies in this paper.\label{tab-nh}}
\setlength{\tabcolsep}{13.5pt}
\tabletypesize{\small}
\tablehead{
  \colhead{Continuum} &
  \colhead{\myt\ Symbol} &
  \colhead{Associated Equivalent} &
  \colhead{\myt\ \xspec\ Table}
  \vspace{-.1cm}
  \\
  \colhead{Type}&
  \colhead{}&
  \colhead{Hydrogen Column Density}&
  \colhead{}
  }
  \colnumbers
  \startdata
  Direct (``zeroth'')               & Z & \nhz\ (line-of-sight)           & {\tt mytorus\_Ezero\_v00.fits} \\
  Compton scattered (``reflected'') & S & \nhs\ (``global'')              & {\tt mytorus\_scatteredH500\_v00.fits} \\
  \enddata
  \renewcommand{\baselinestretch}{0.5} 
  \tablecomments{\nhs\ is also the column density associated with \feka\ and
    $\beta$ fluorescent line emission. The direct continuum is only attenuated along the line-of-sight by the
    associated column density \nhz\ and
    is not affected by Compton scattering. In contrast, the reflected continuum is Compton scattered out of the
    line-of-sight by the associated column density \nhs. A column density \nhinter\ due to matter between the observer and the source
    further attenuates the
    total continuum reaching the observer. See also \fr{fig-NHplot}.}
  \renewcommand{\baselinestretch}{1} 
  \vspace{-1cm}
  \end{deluxetable*}

%% file: fig-NuSTAR.tex
\begin{figure*}[t!]
    \begin{minipage}[c]{0.45\textwidth}\hspace{-.7cm}
        \includegraphics[width=0.84\textwidth,angle=-90]{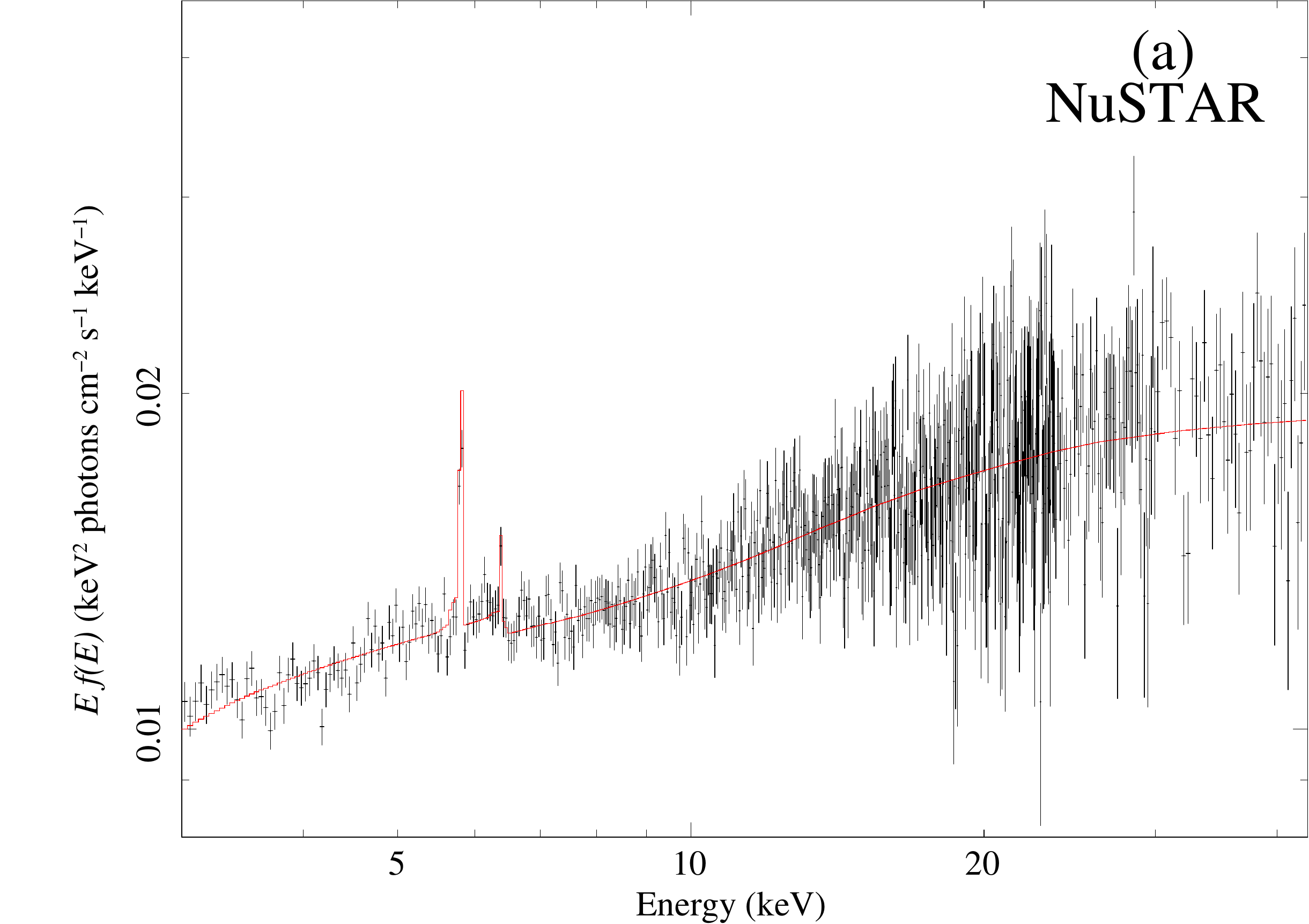}
    \end{minipage}
    \hspace{0.5cm}
    \begin{minipage}[c]{0.45\textwidth}\vspace{0pt}
        \includegraphics[width=0.84\textwidth,angle=-90]{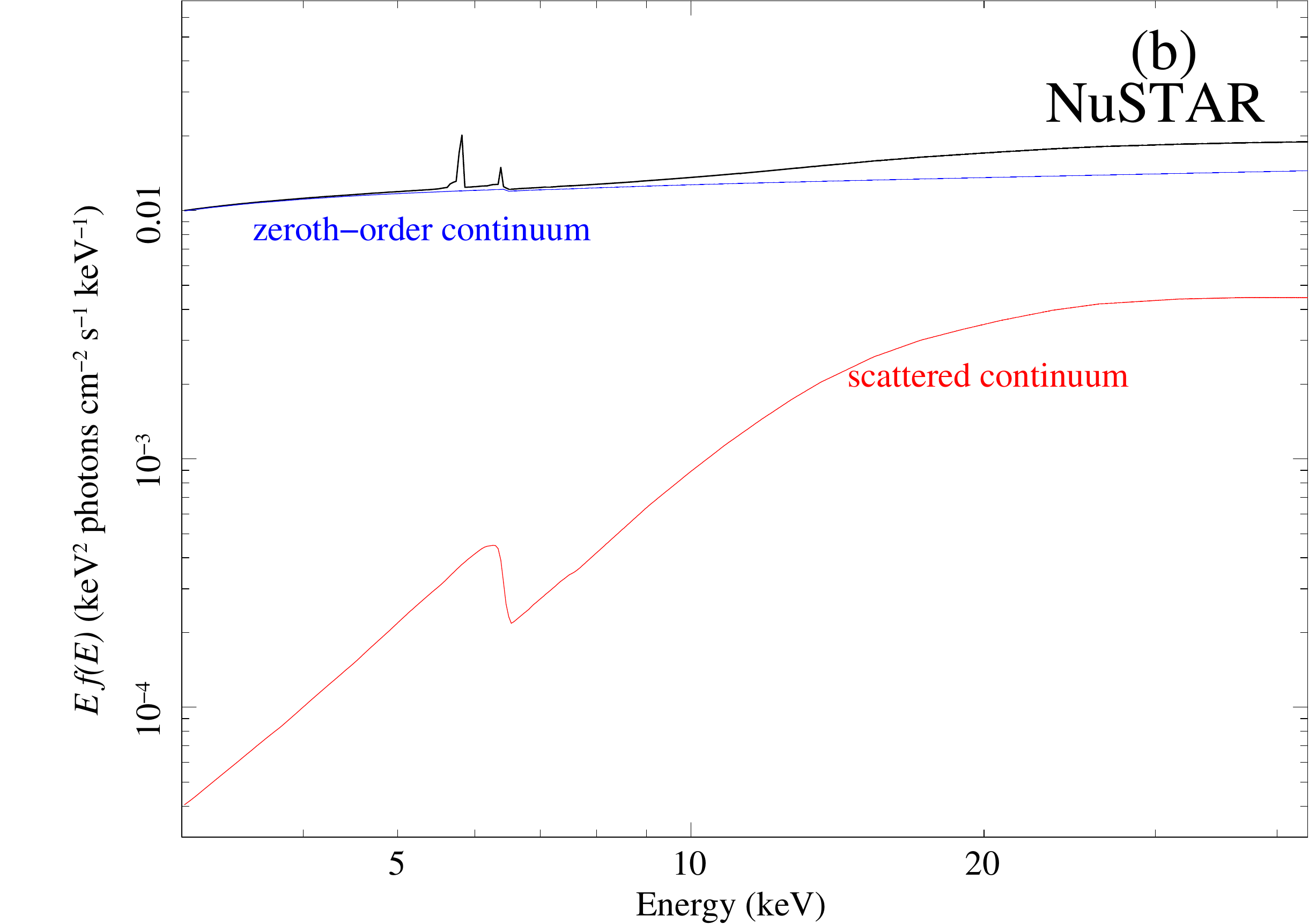}
    \end{minipage}\\

    \begin{minipage}[c]{0.45\textwidth}\hspace{-.65cm}
        \includegraphics[width=0.84\textwidth,angle=-90]{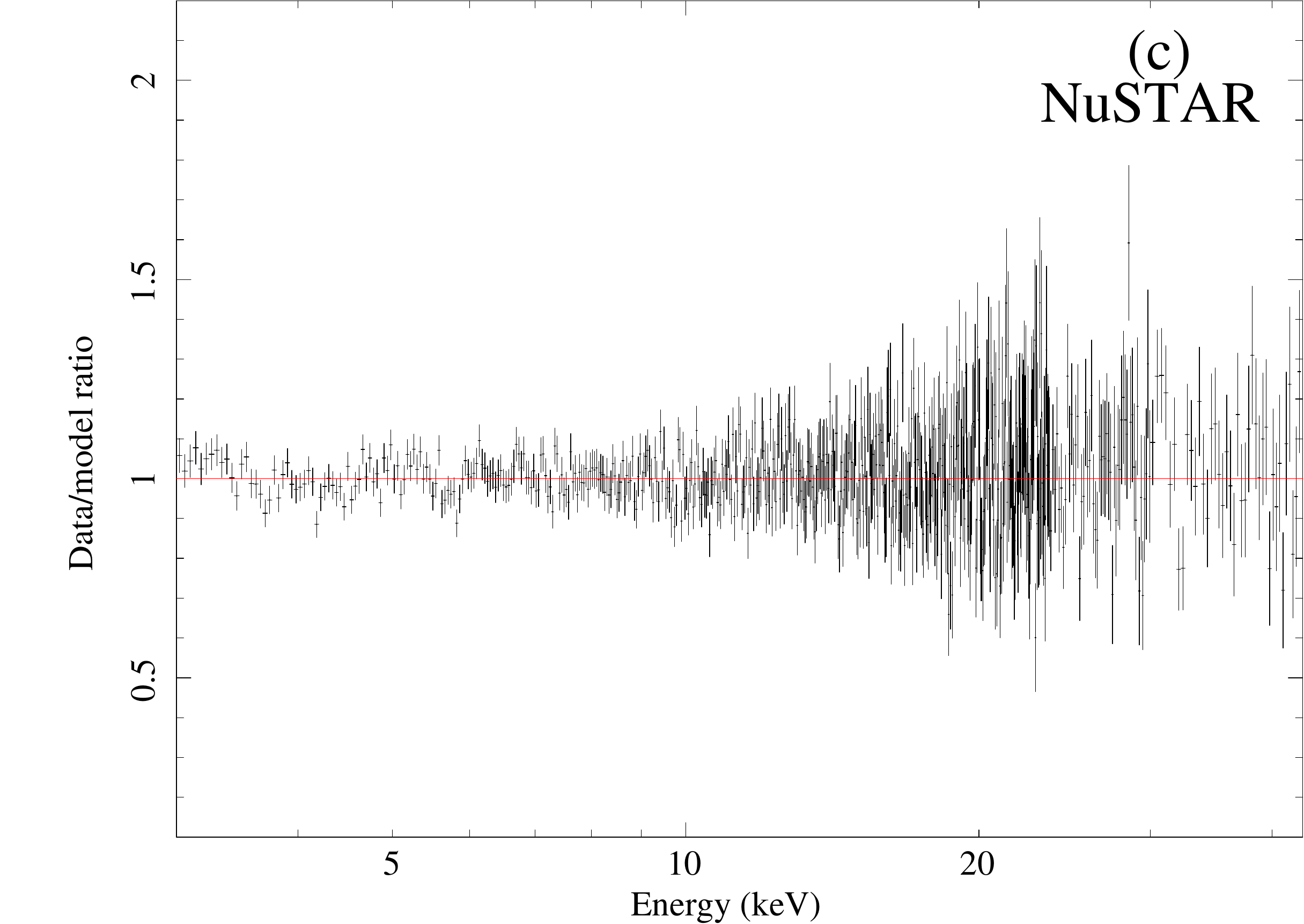}
    \end{minipage}
    \hspace{0.52cm}
    \begin{minipage}[c]{0.45\textwidth}\vspace{-30pt}
        \includegraphics[width=0.98\textwidth,angle=-90]{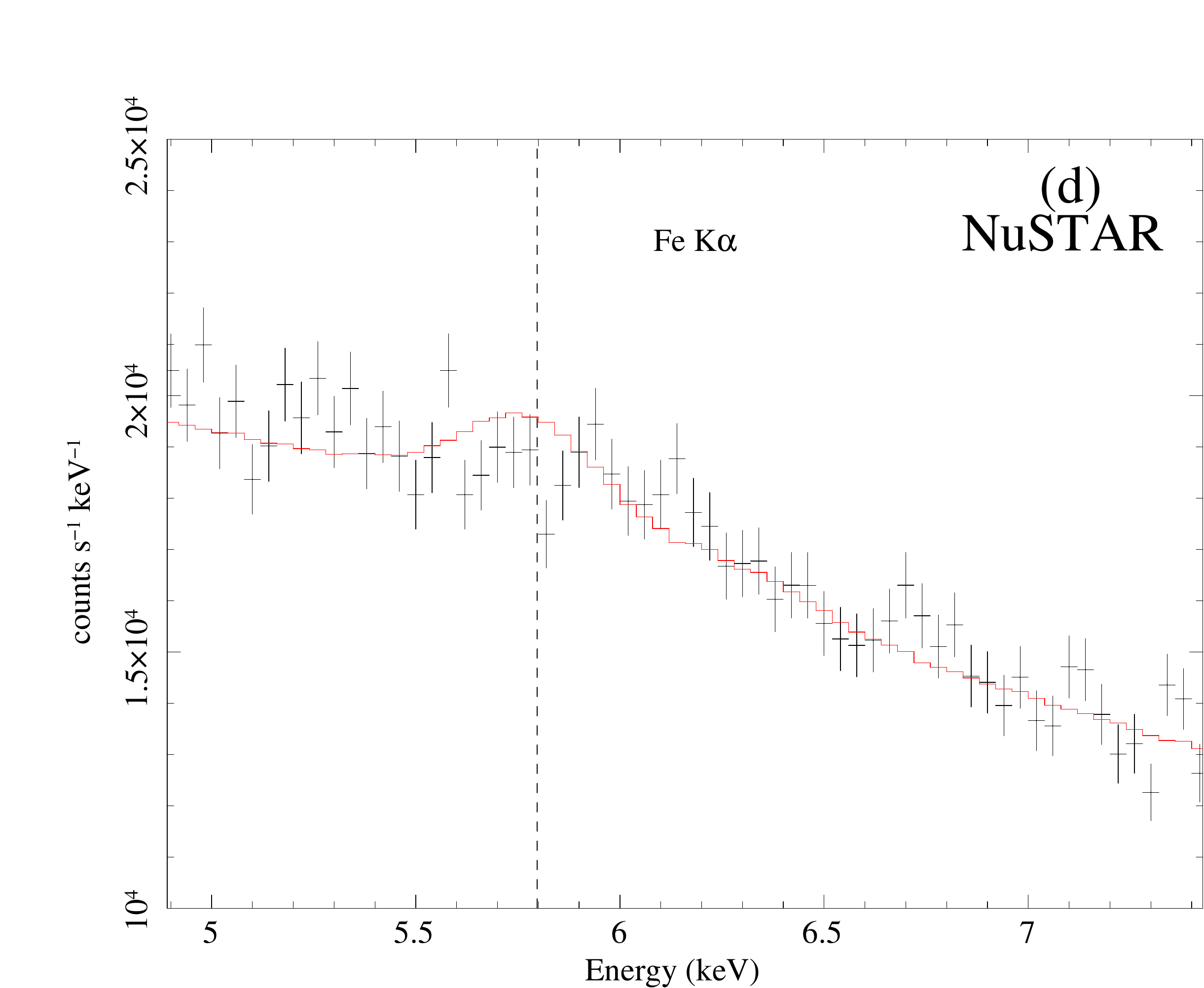}
    \end{minipage}
    \vspace{-0cm}
  \caption{\footnotesize Results of spectral fitting of the co-added \nustar\ observations with \myt. FPMA and FPMB results are combined for plotting purposes only. Panels are (a)-data and total model over the full spectral range fitted, (b)-total model ({\it black}) and continuum components, (c)-data/model ratio, and (d)-data and total model in the vicinity of the \feka\ emission line, respectively.
  \label{fig-NuSTAR}}
\end{figure*}

%% file: fig-702-xis.tex
\begin{figure*}[t!]
    \begin{minipage}[c]{0.45\textwidth}\hspace{-1cm}
        \includegraphics[width=0.84\textwidth,angle=-90]{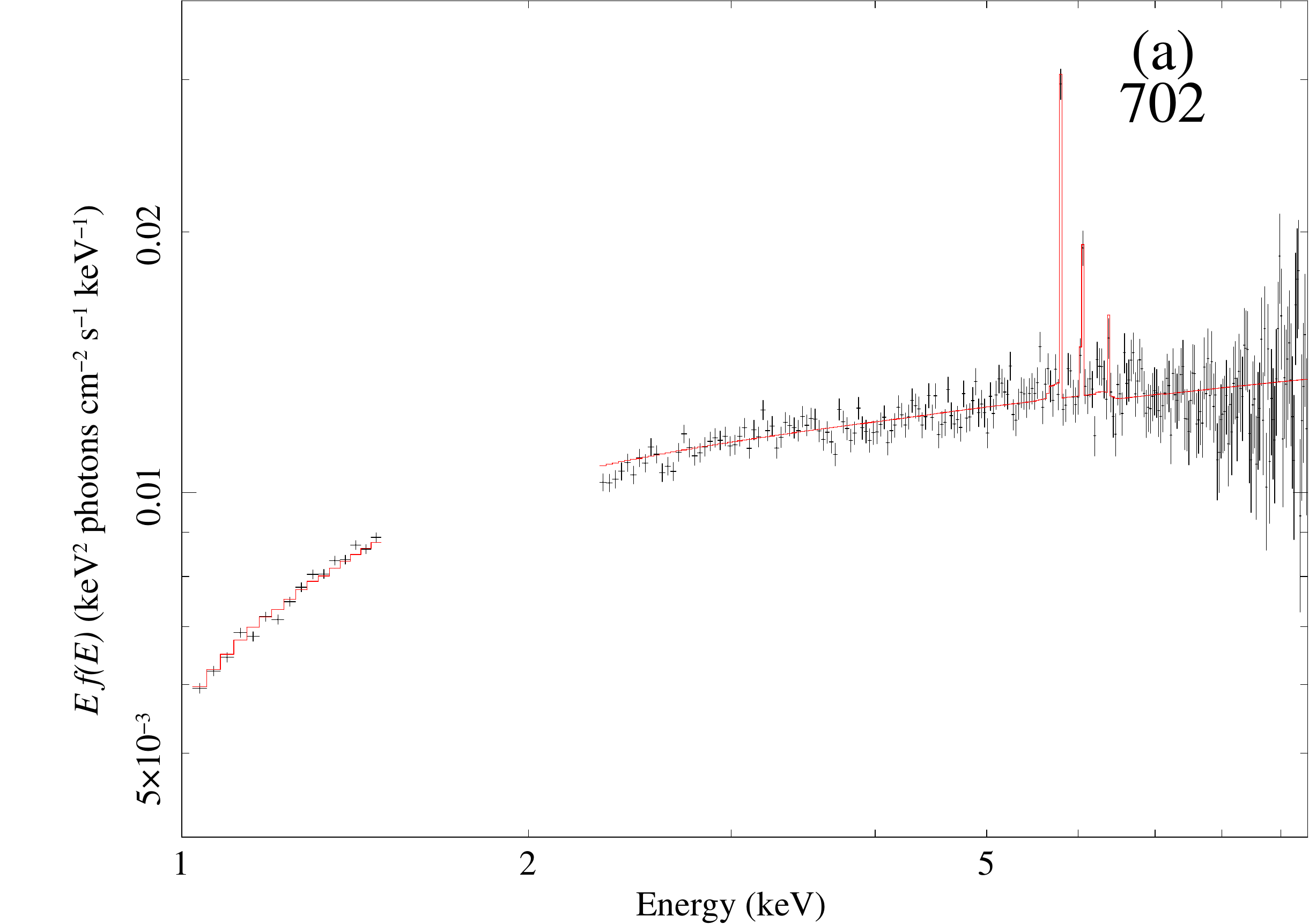}
    \end{minipage}
    \hspace{0.5cm}
    \begin{minipage}[c]{0.45\textwidth}\vspace{0pt}
        \includegraphics[width=0.84\textwidth,angle=-90]{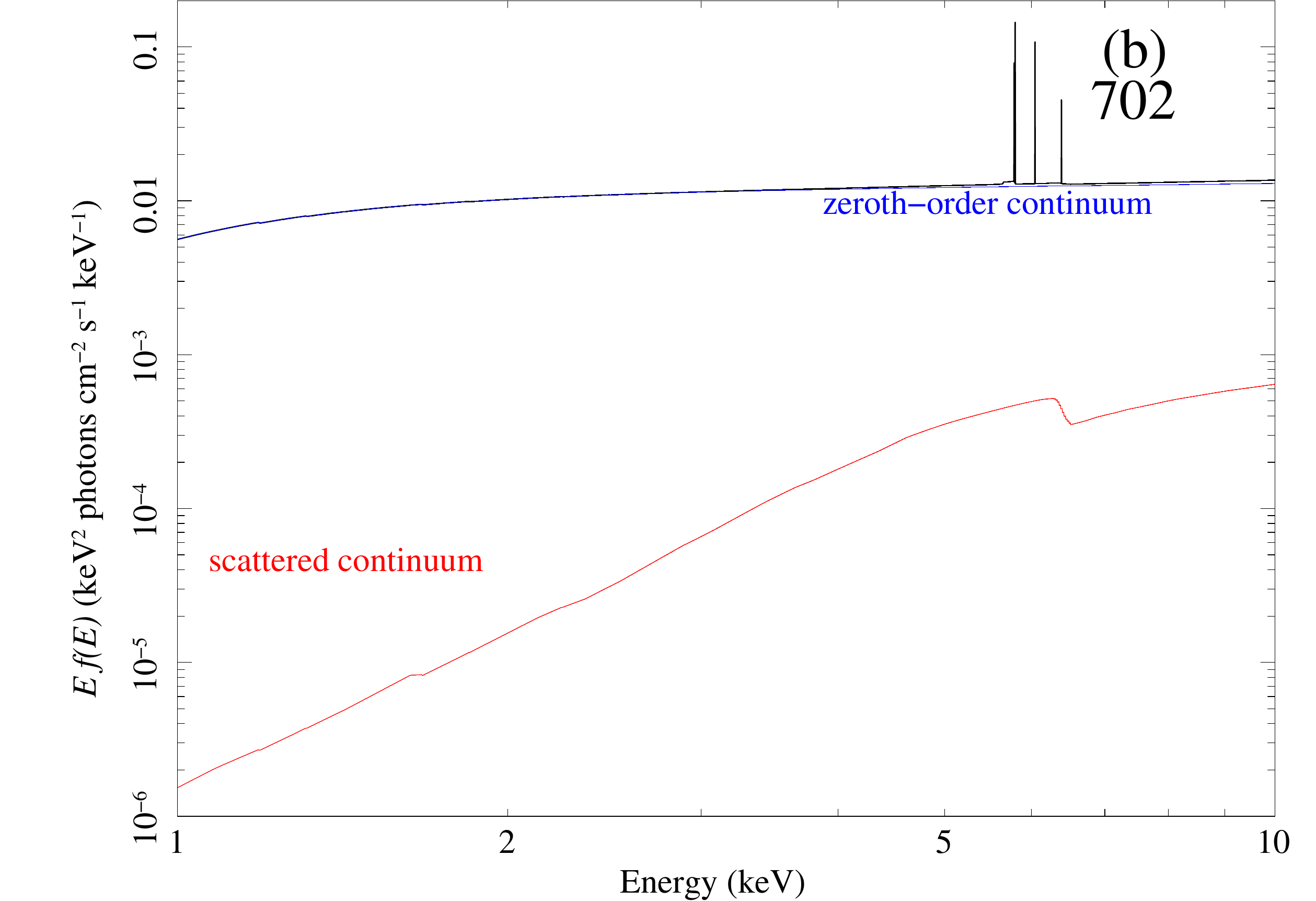}
    \end{minipage}\\

    \begin{minipage}[c]{0.45\textwidth}\hspace{-.95cm}
        \includegraphics[width=0.865\textwidth,angle=-90]{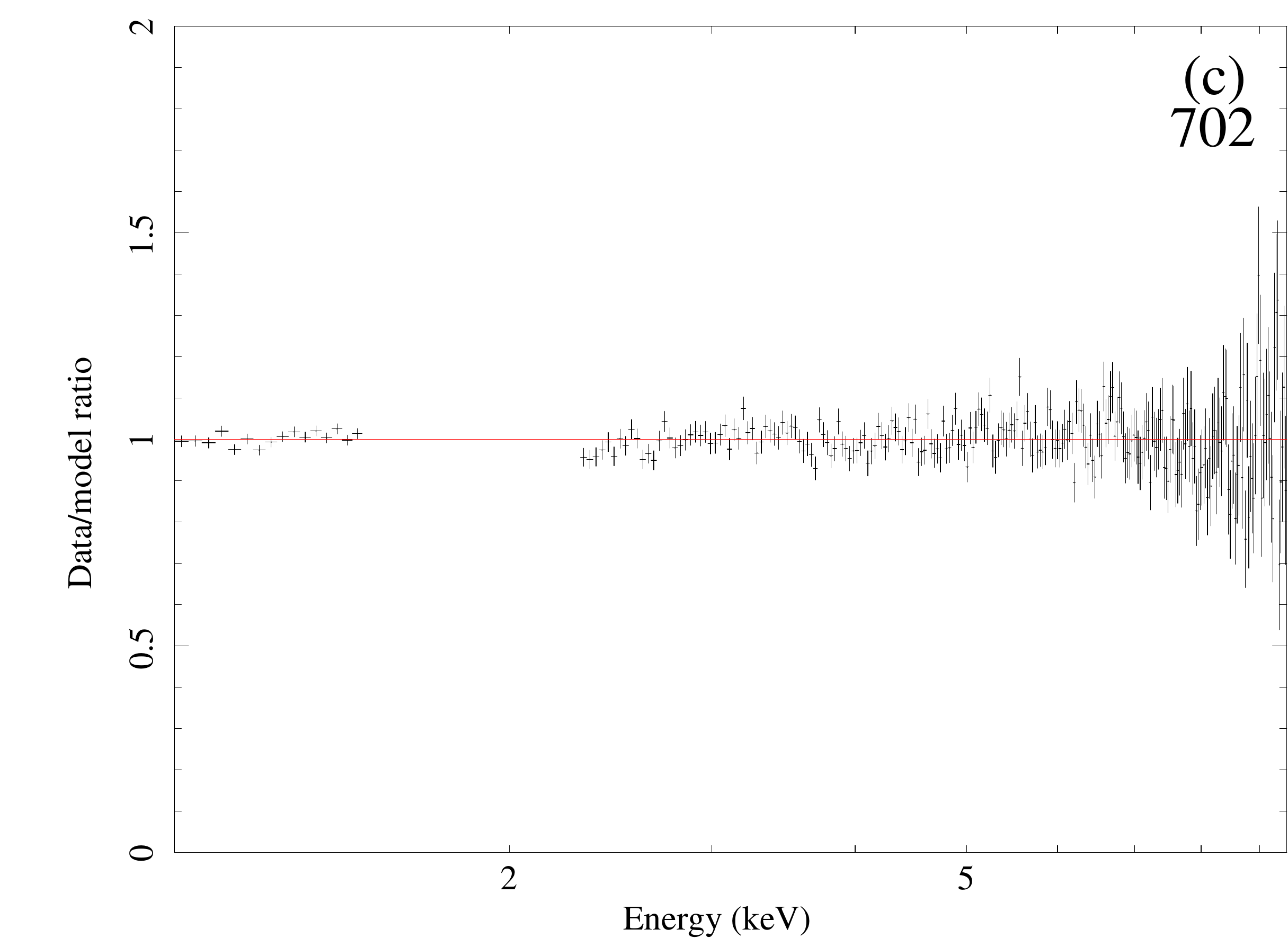}
    \end{minipage}
    \hspace{0.52cm}
    \begin{minipage}[c]{0.45\textwidth}\vspace{-28pt}
        \includegraphics[width=0.97\textwidth,angle=-90]{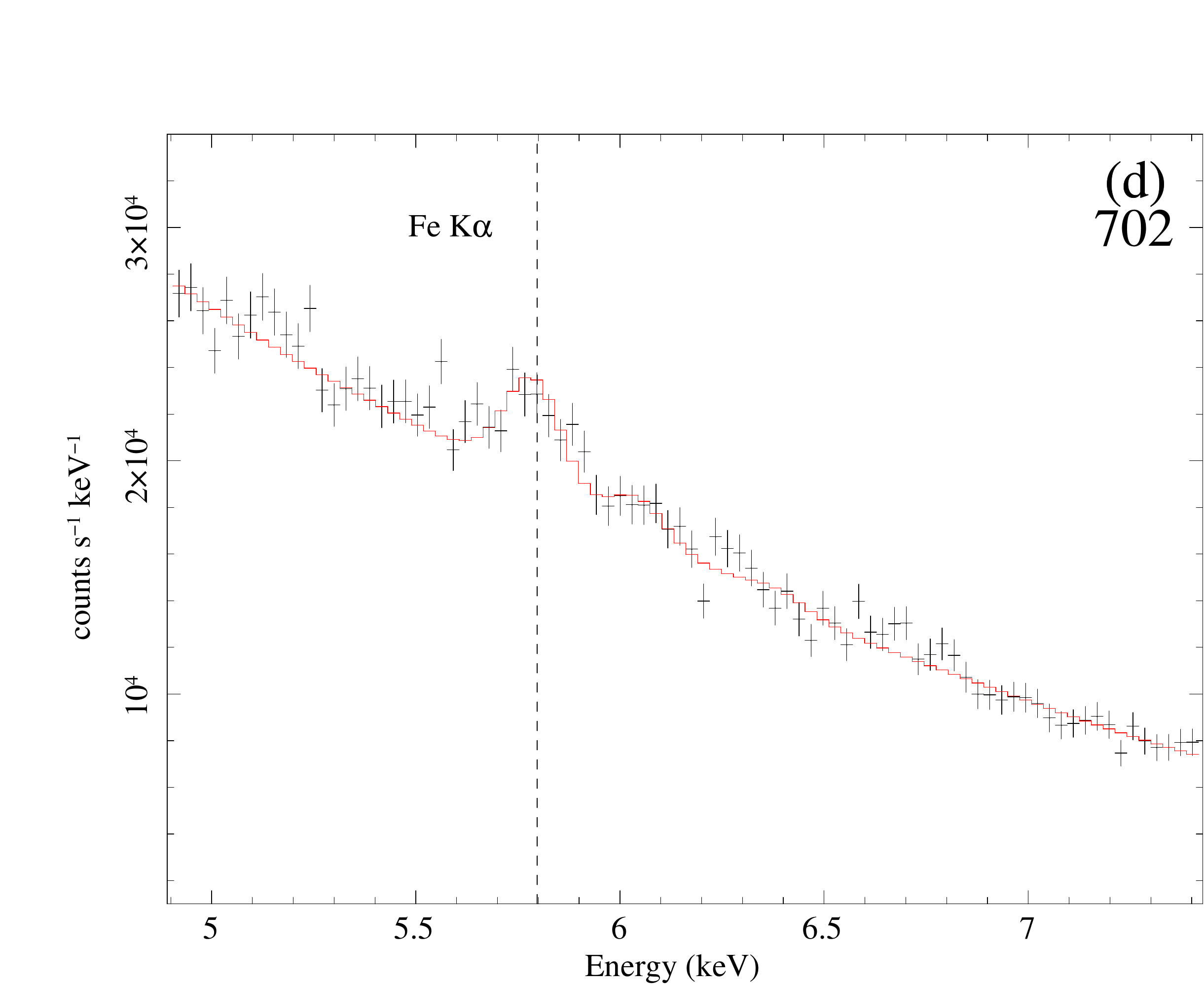}
    \end{minipage}
    \vspace{-0cm}
  \caption{\footnotesize Results of spectral fitting to \suzaku-XIS obs.~702 with \myt. Panels are as in \fr{fig-NuSTAR}.
  \label{fig-702-xis}}
\end{figure*}
%
%\begin{figure*}[t!]
%  \gridline{\fig{4C7426_702_myun_XIS6_DA.pdf}{0.35\textwidth}{}
%            \fig{4C7426_702_myun_XIS6_RA.pdf}{0.35\textwidth}{}
%  }
%  \gridline{\fig{4C7426_702_myun_XIS6_MO.pdf}{0.35\textwidth}{}
%    \fig{4C7426_702_myun_XIS6_Fe.pdf}{0.35\textwidth}{}
%  }
%  \caption{Results of spectral fitting to \suzaku-XIS obs.~702 with \myt.
%  \label{fig-702}}
%\end{figure*}
%  

%% file: fig-702-xispin.tex
\begin{figure*}[t!]
    \begin{minipage}[c]{0.45\textwidth}\hspace{-.4cm}
        \includegraphics[width=0.8\textwidth,angle=-90]{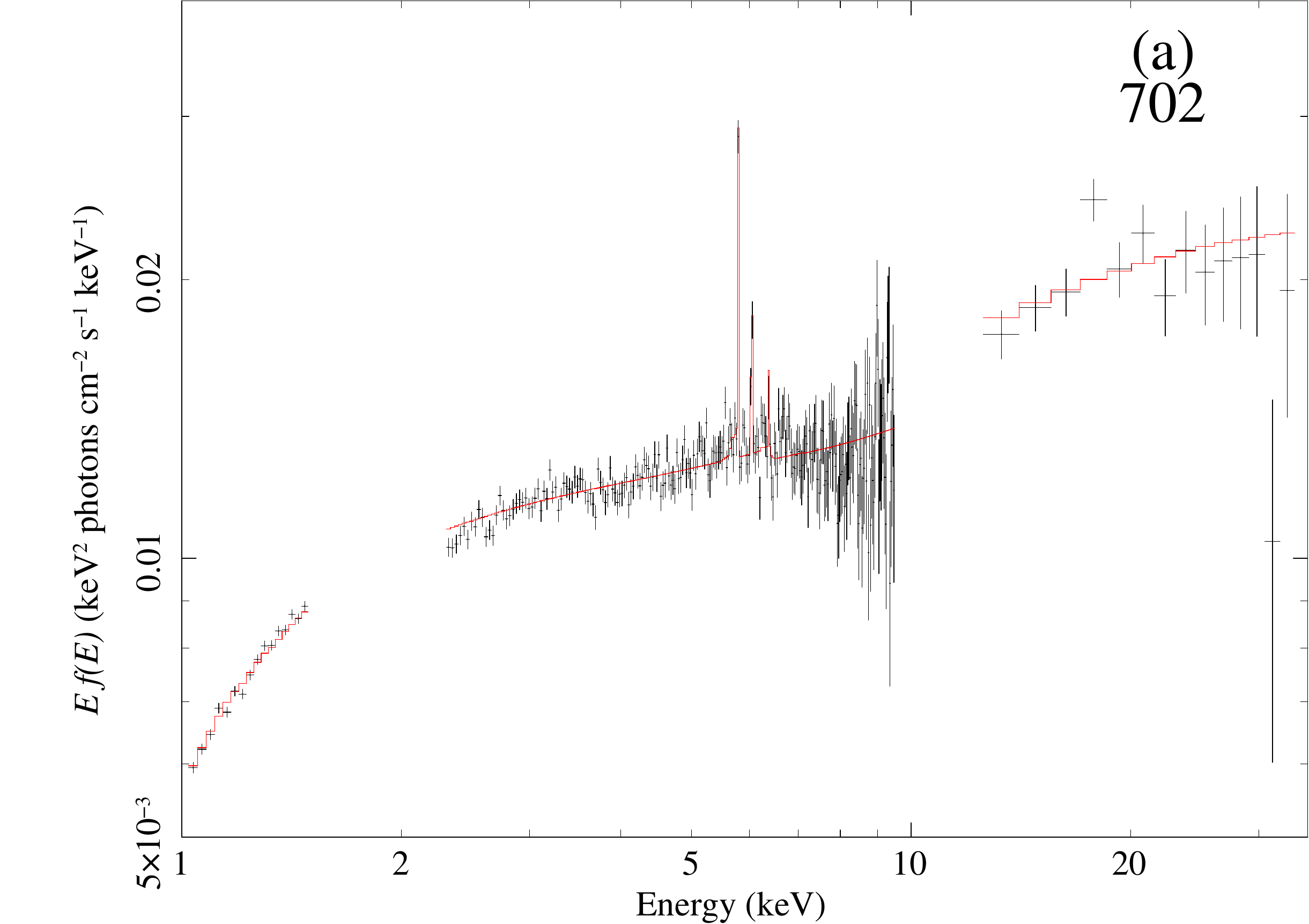}
    \end{minipage}
    \hspace{0.5cm}
    \begin{minipage}[c]{0.45\textwidth}\vspace{-6pt}
        \includegraphics[width=0.825\textwidth,angle=-90]{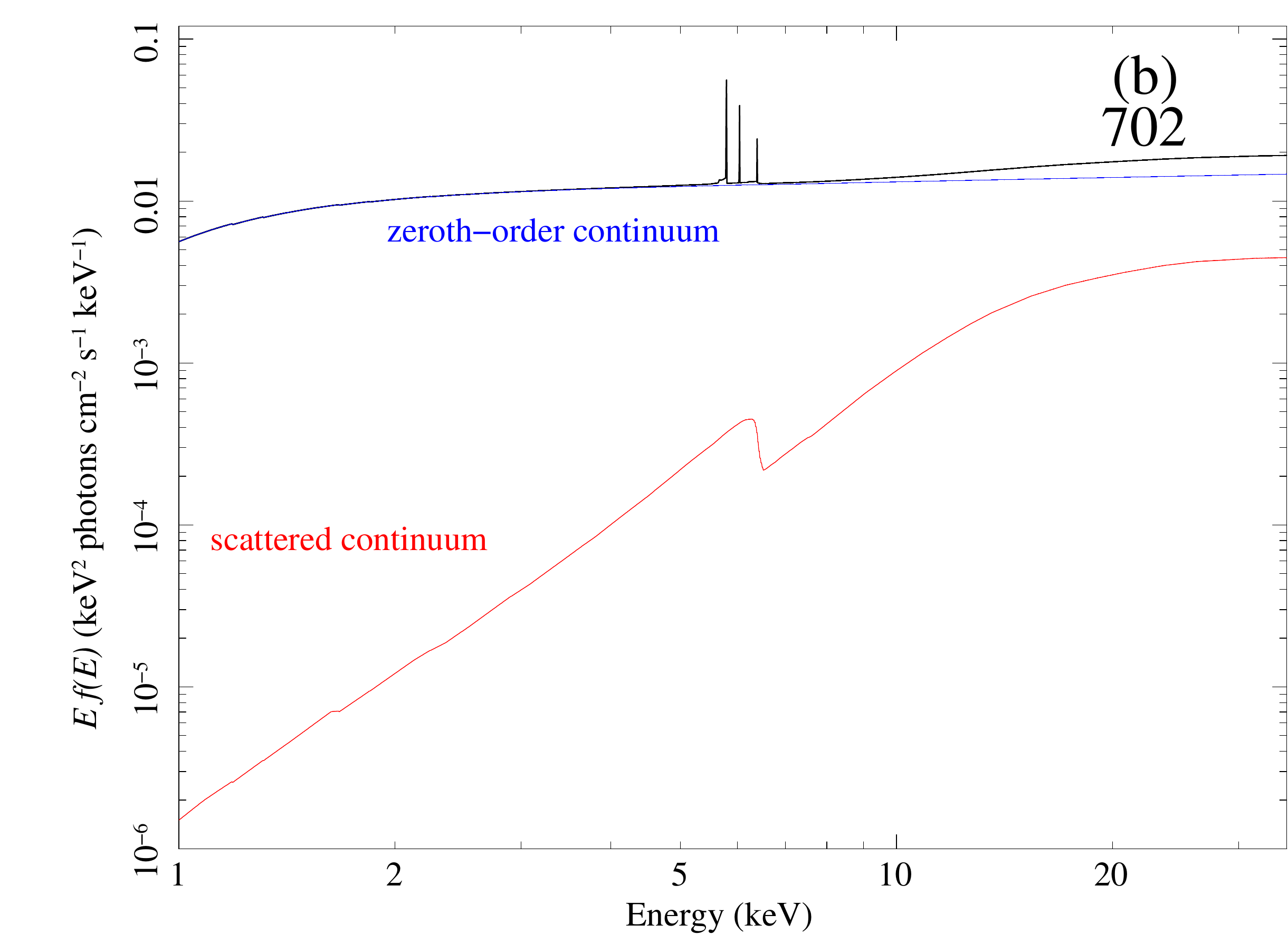}
    \end{minipage}\\

    \begin{minipage}[c]{0.45\textwidth}\hspace{-.4cm}
        \includegraphics[width=0.825\textwidth,angle=-90]{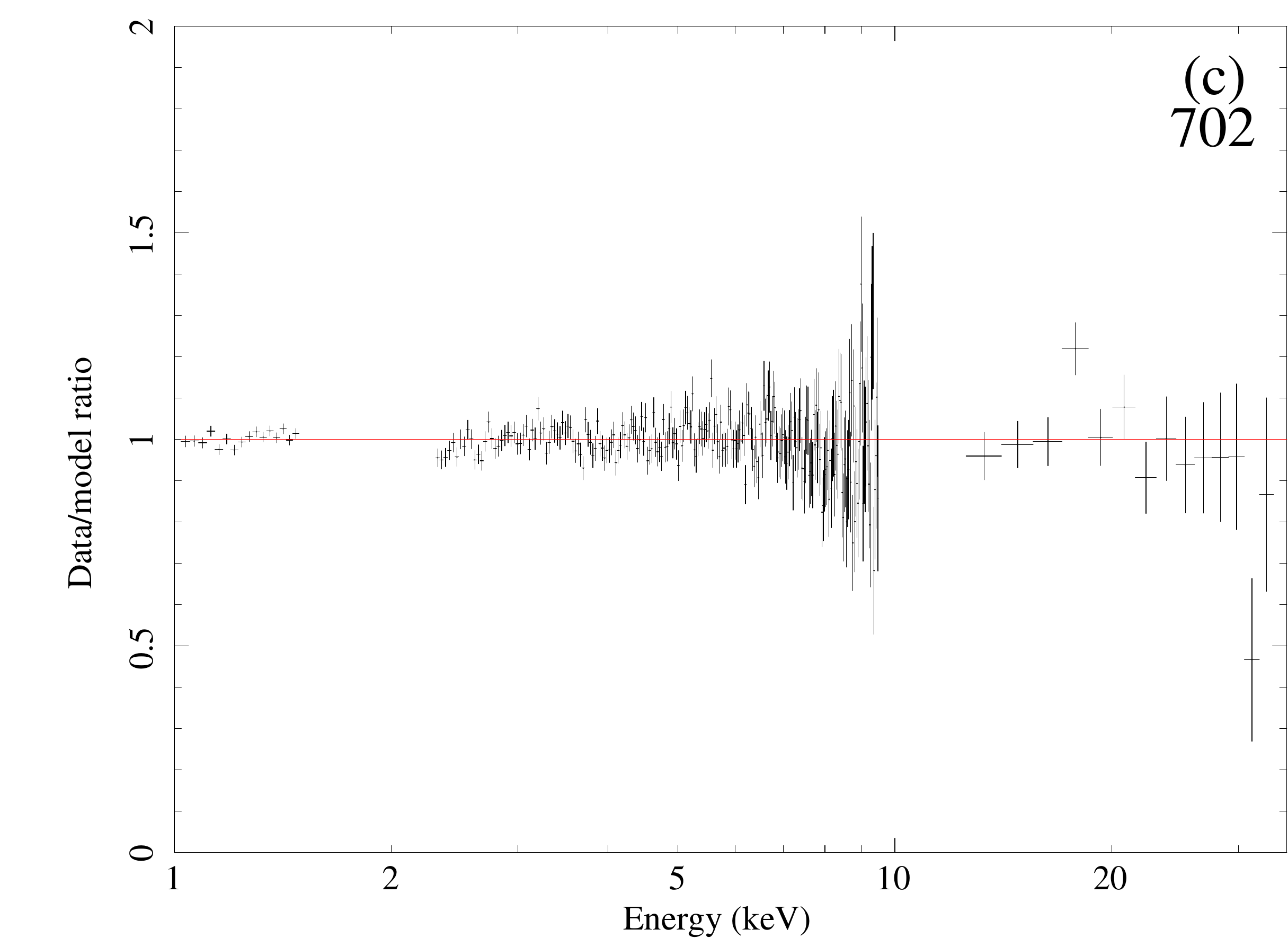}
    \end{minipage}
    \hspace{0.52cm}
    \begin{minipage}[c]{0.45\textwidth}\vspace{-22pt}
        \includegraphics[width=0.91\textwidth,angle=-90]{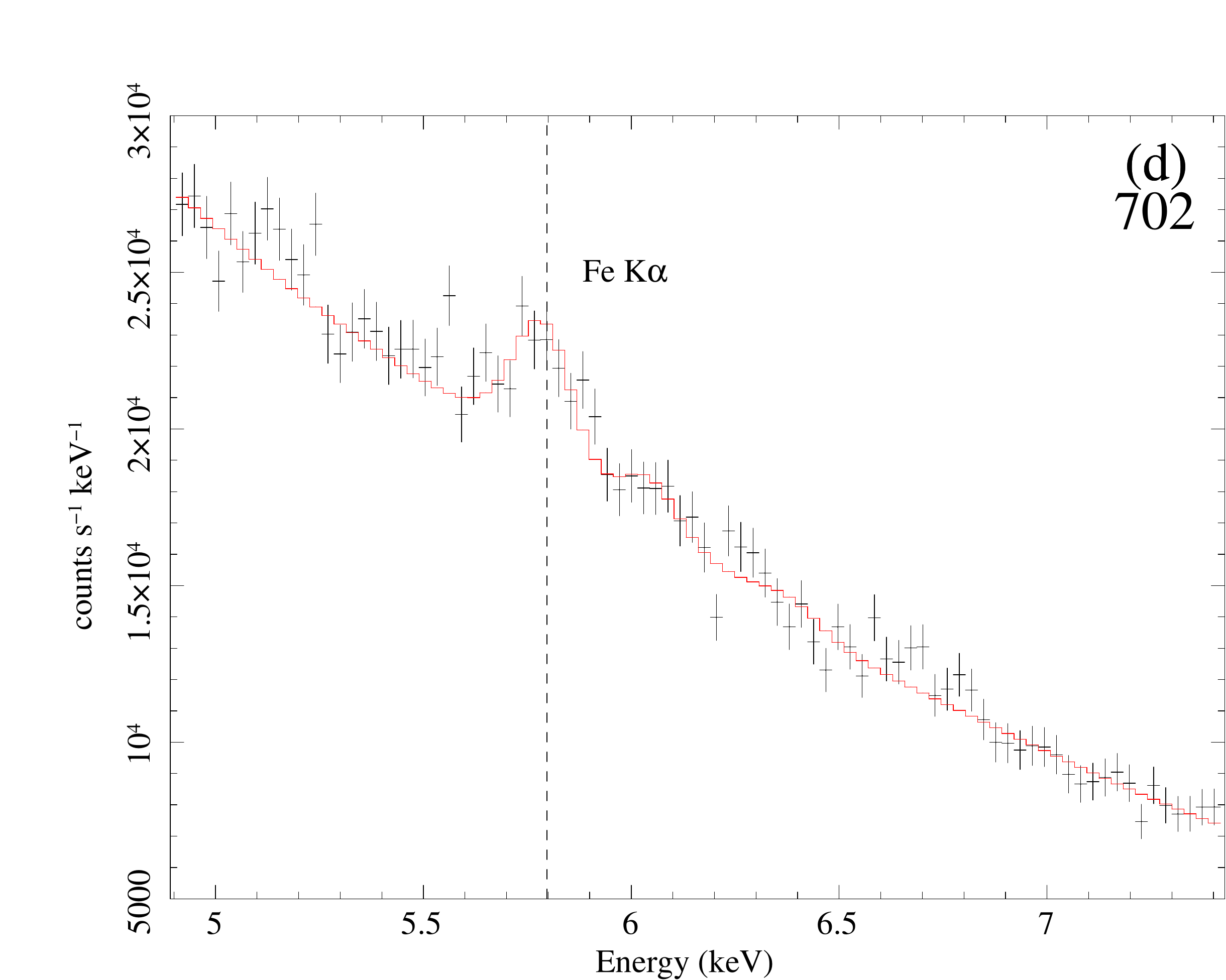}
    \end{minipage}
    \vspace{-0cm}
  \caption{\footnotesize Results of spectral fitting to \suzaku-XIS+PIN obs.~702 with \myt. Panels are as in \fr{fig-702-xis}.
  \label{fig-702-xispin}}
\end{figure*}

%% file: fig-706-xis.tex
\begin{figure*}[t!]
    \begin{minipage}[c]{0.45\textwidth}\hspace{-1cm}
        \includegraphics[width=0.84\textwidth,angle=-90]{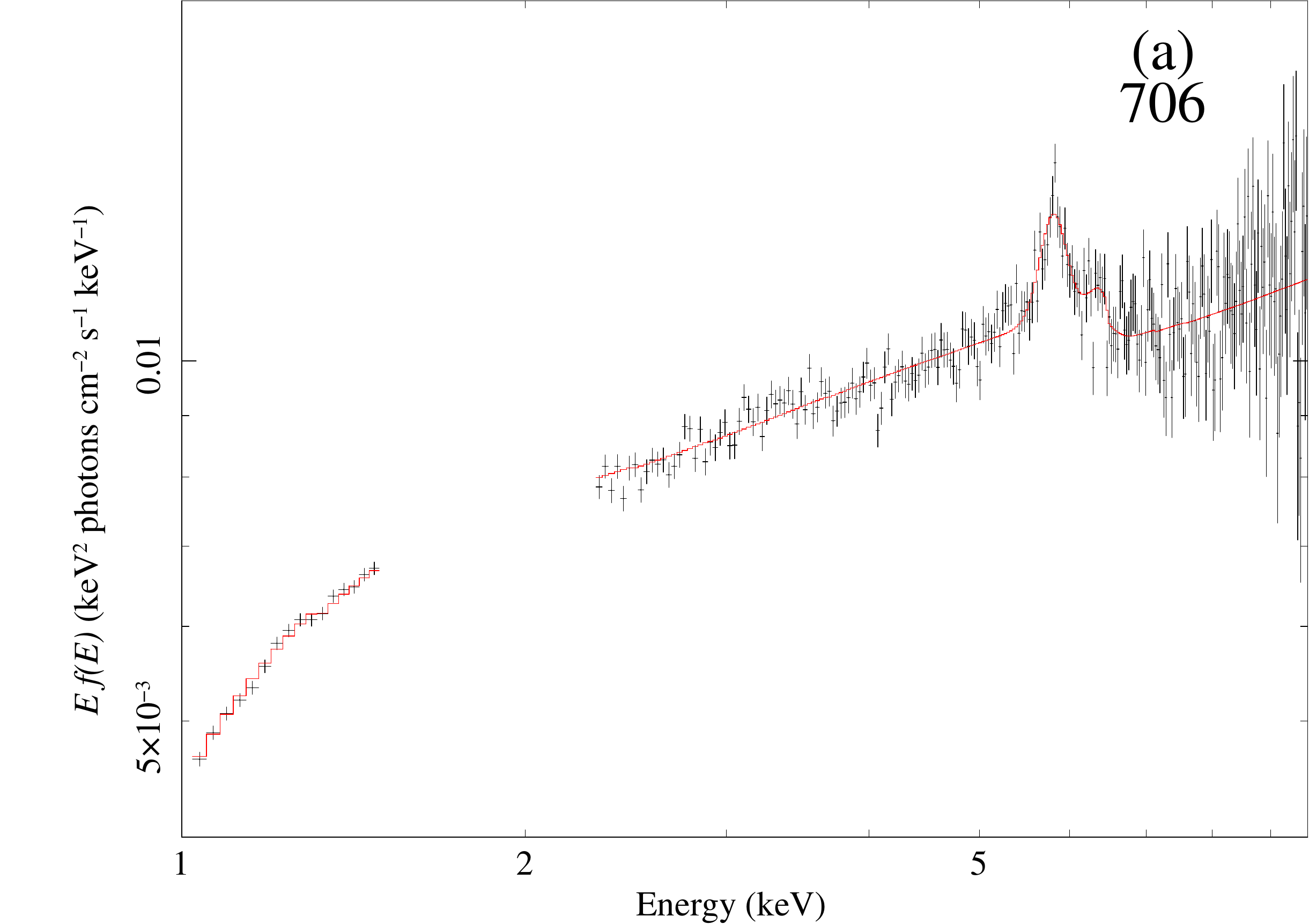}
    \end{minipage}
    \hspace{0.5cm}
    \begin{minipage}[c]{0.45\textwidth}\vspace{-10pt}
        \includegraphics[width=0.87\textwidth,angle=-90]{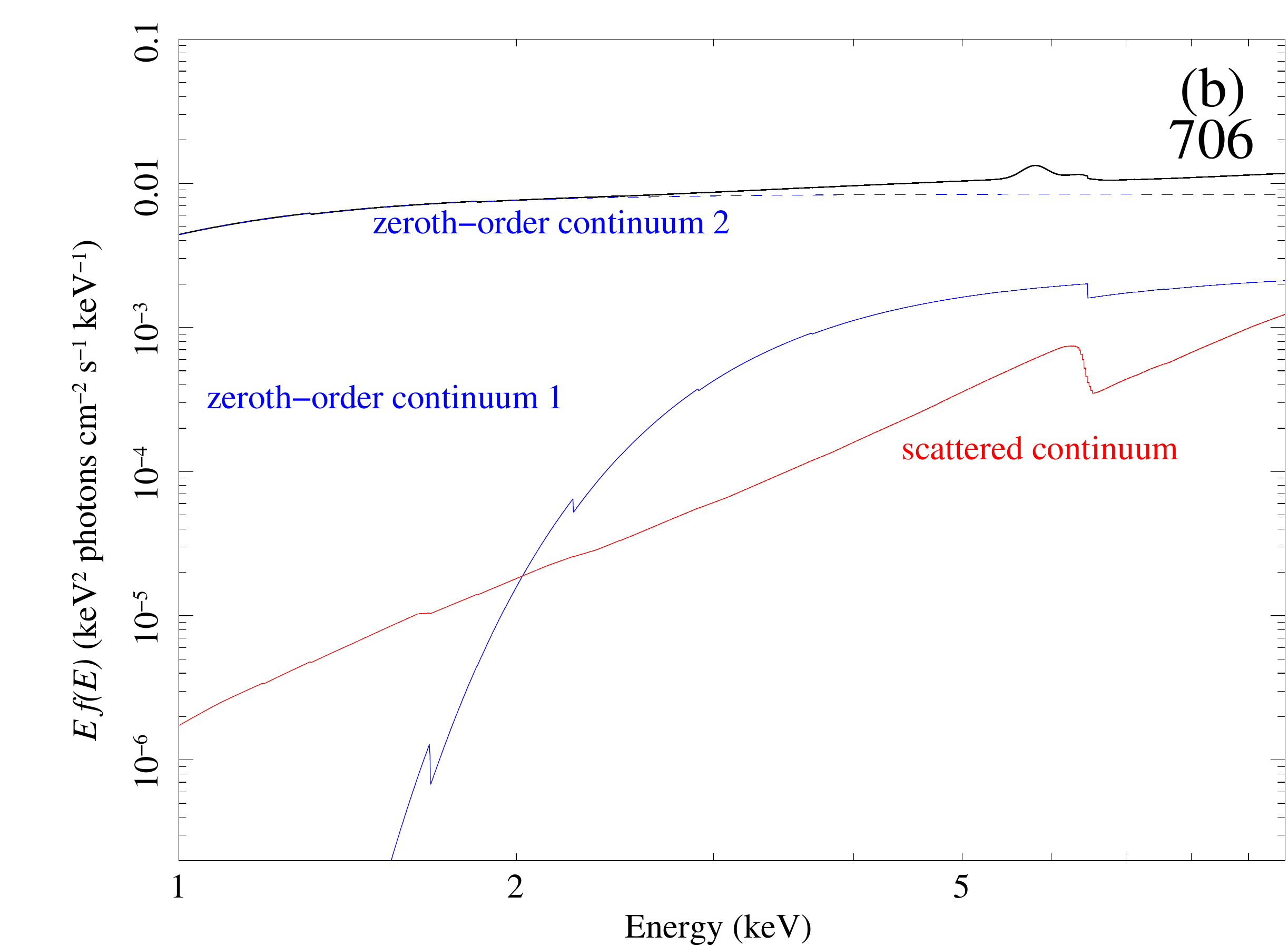}
    \end{minipage}\\

    \begin{minipage}[c]{0.45\textwidth}\hspace{-1cm}
        \includegraphics[width=0.87\textwidth,angle=-90]{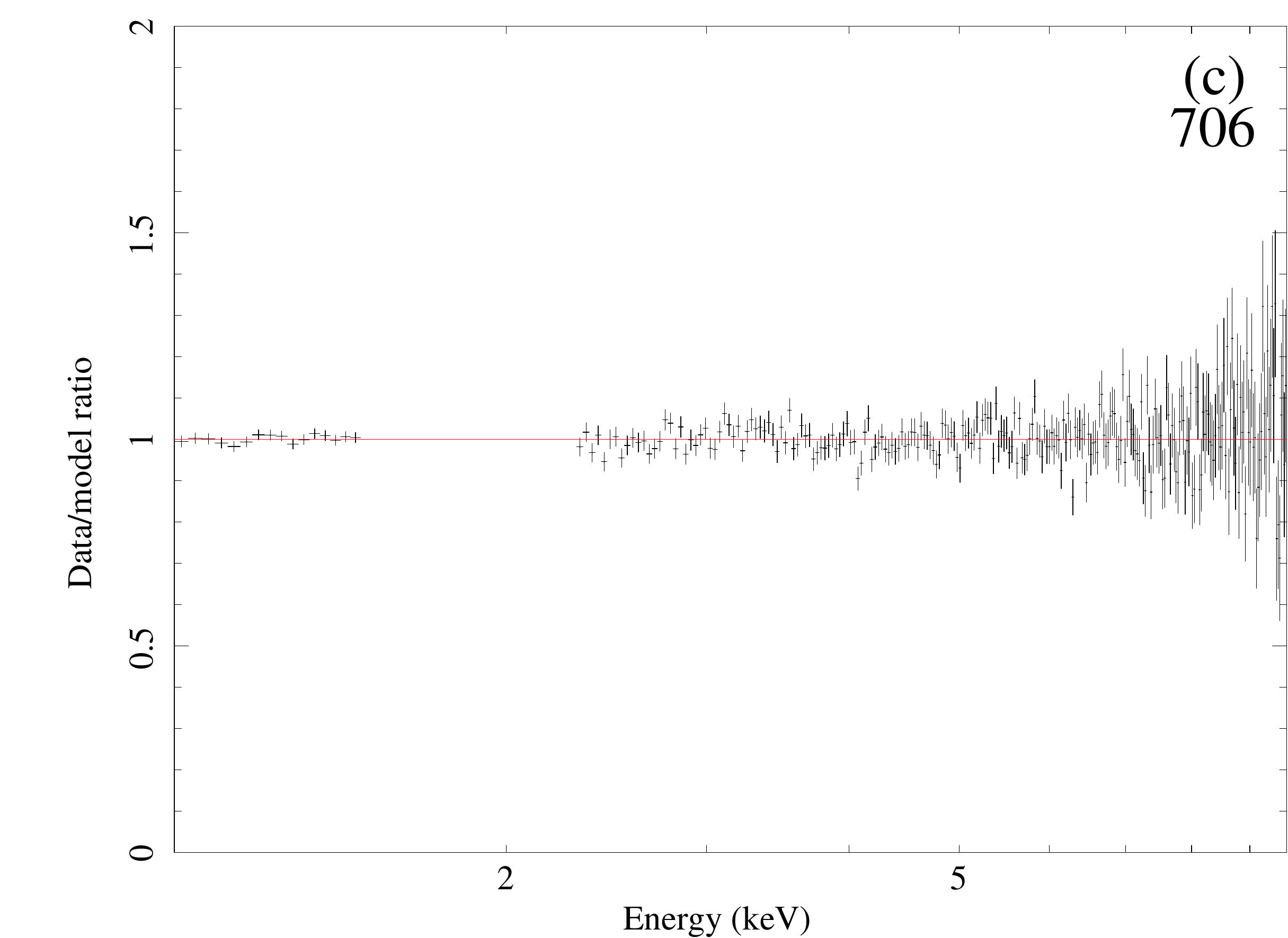}
    \end{minipage}
    \hspace{0.5cm}
    \begin{minipage}[c]{0.45\textwidth}\vspace{-30pt}
        \includegraphics[width=0.99\textwidth,angle=-90]{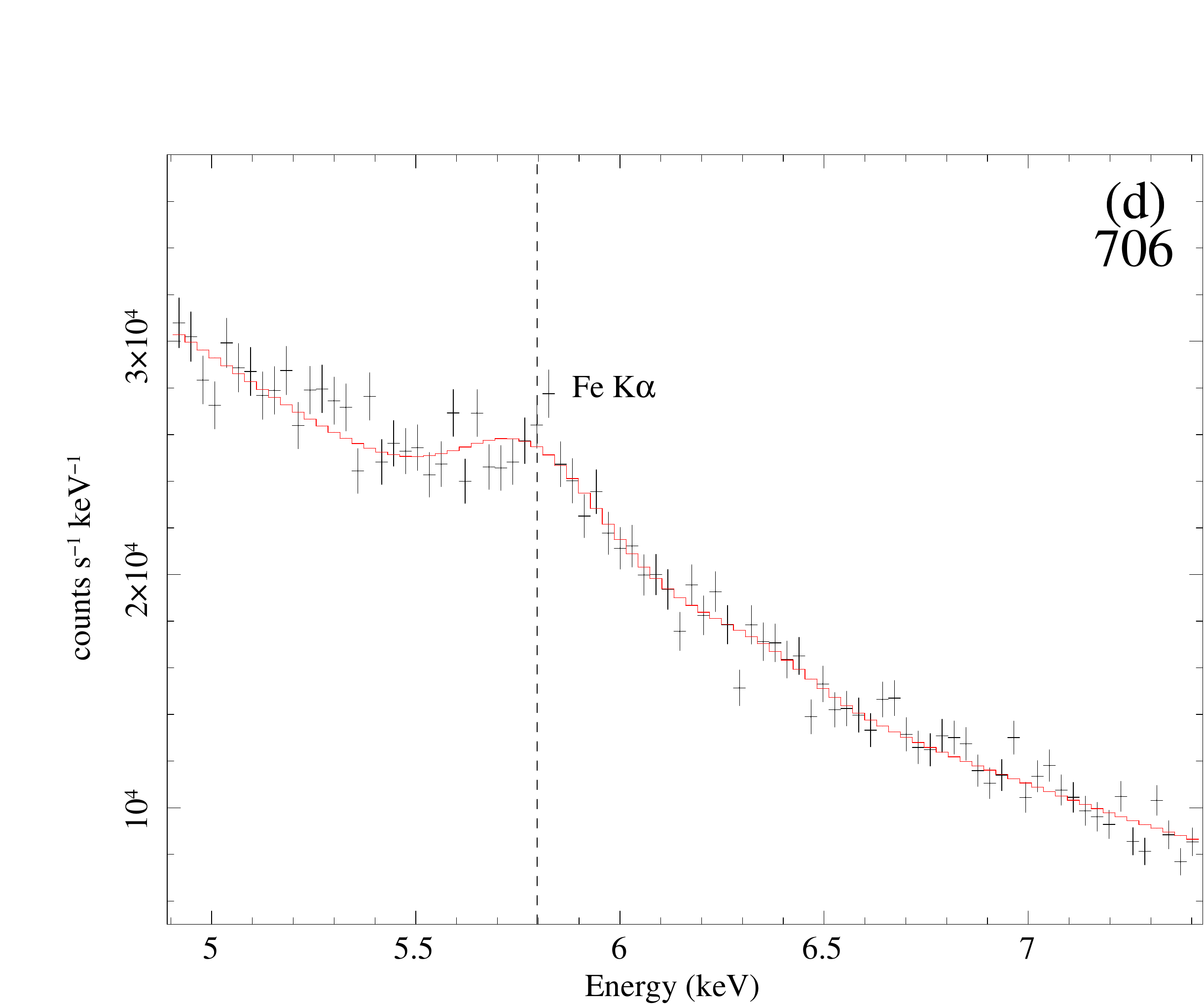}
    \end{minipage}
    \vspace{-0cm}
  \caption{\footnotesize Results of spectral fitting to \suzaku-XIS obs.~706 with \myt. Panels are as in \fr{fig-702-xis}.
  \label{fig-706-xis}}
\end{figure*}

%% file: fig-706-xispin.tex
\begin{figure*}[t!]
    \begin{minipage}[c]{0.45\textwidth}\hspace{-.4cm}
        \includegraphics[width=0.8\textwidth,angle=-90]{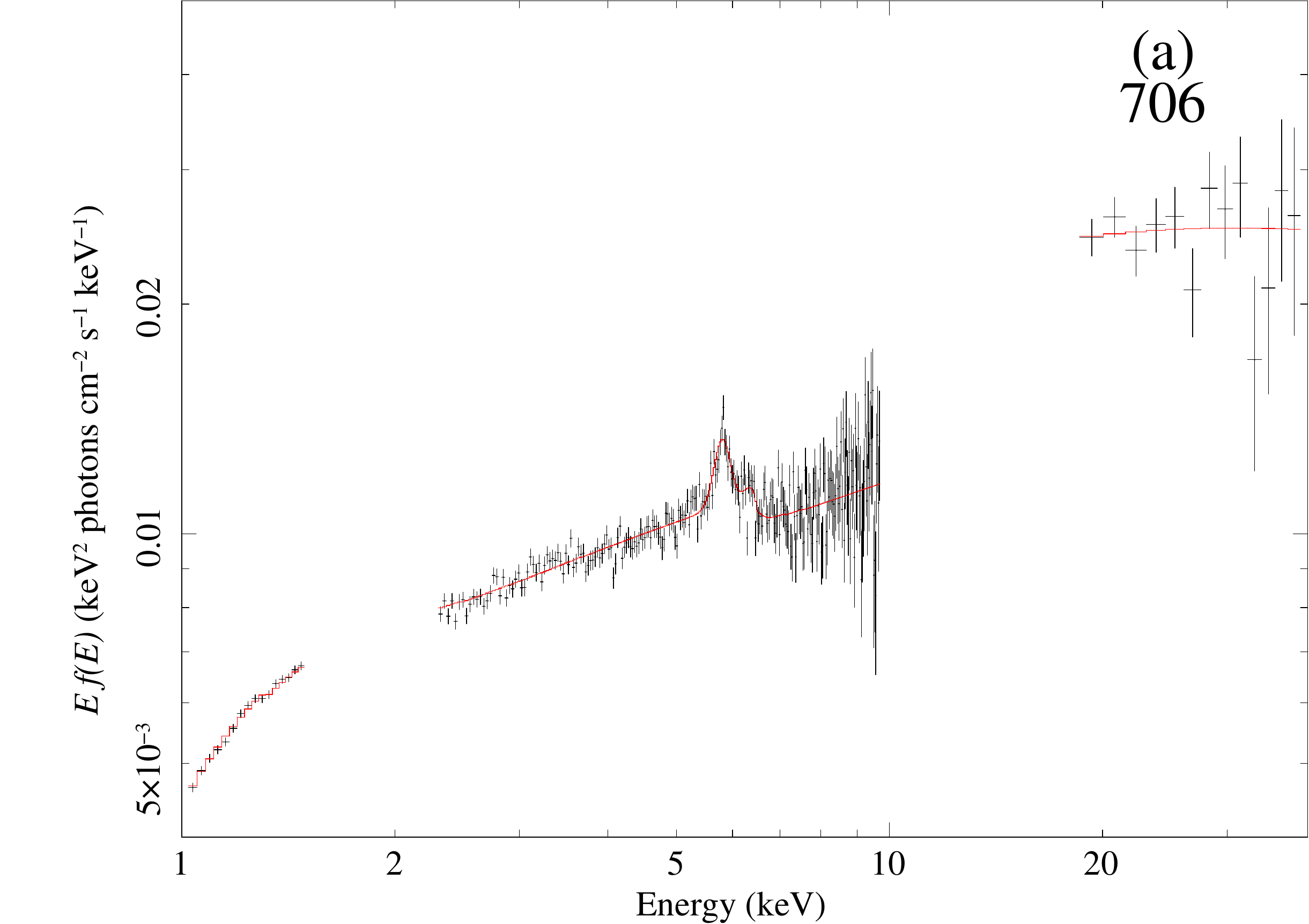}
    \end{minipage}
    \hspace{0.5cm}
    \begin{minipage}[c]{0.45\textwidth}\vspace{0pt}
        \includegraphics[width=0.8\textwidth,angle=-90]{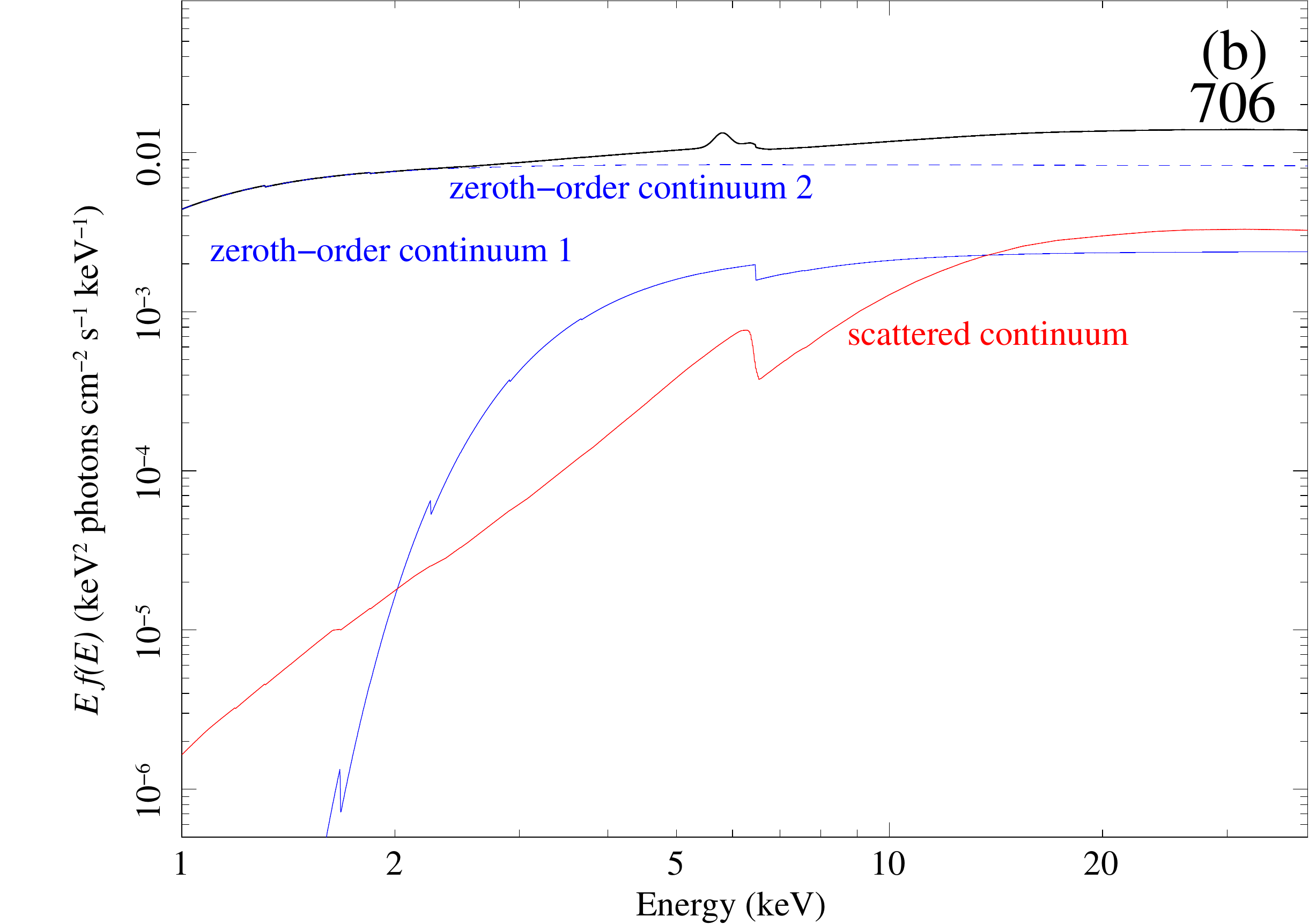}
    \end{minipage}\\

    \begin{minipage}[c]{0.45\textwidth}\hspace{-.4cm}
        \includegraphics[width=0.83\textwidth,angle=-90]{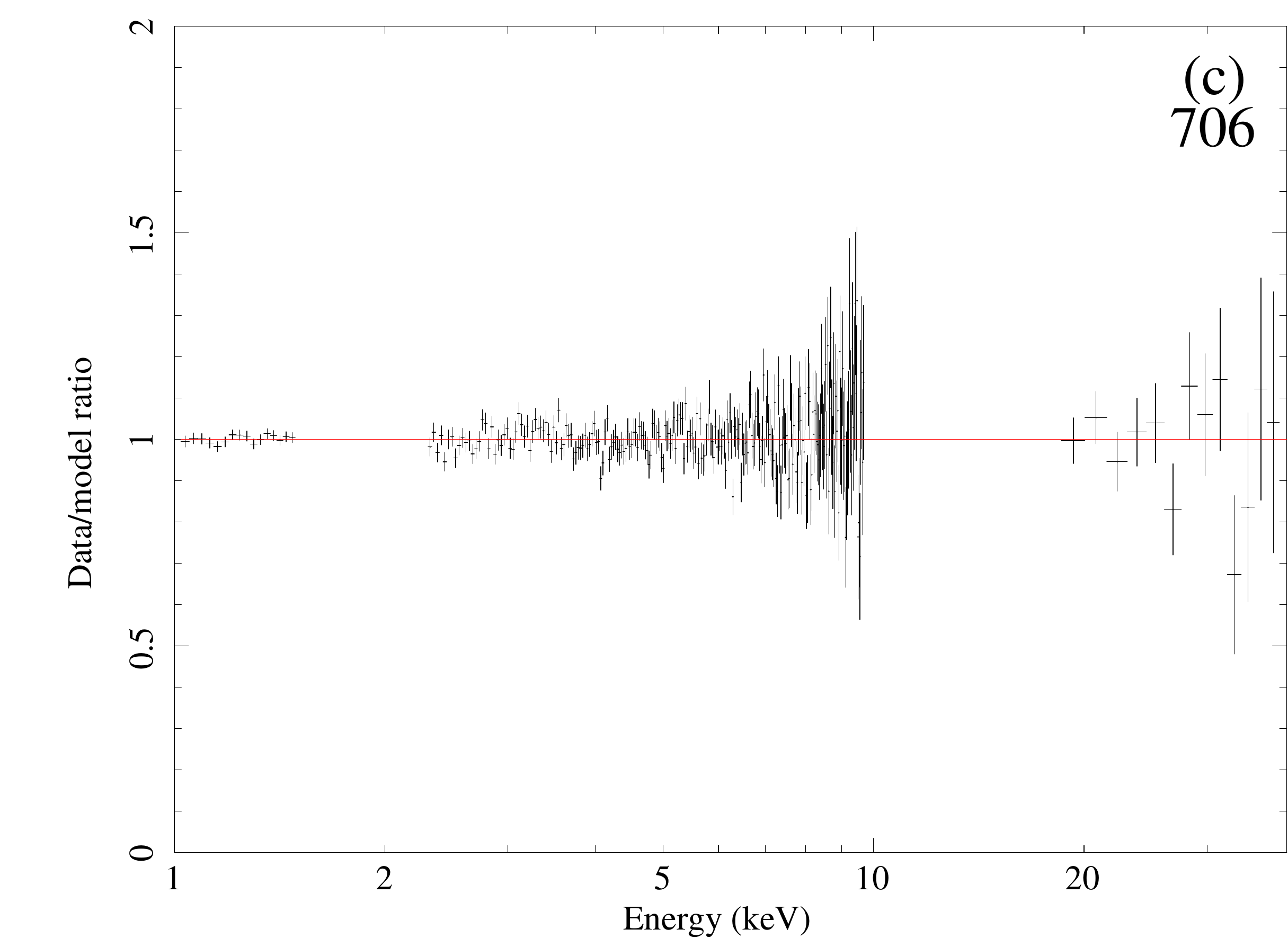}
    \end{minipage}
    \hspace{0.48cm}
    \begin{minipage}[c]{0.45\textwidth}\vspace{-24pt}
        \includegraphics[width=0.94\textwidth,angle=-90]{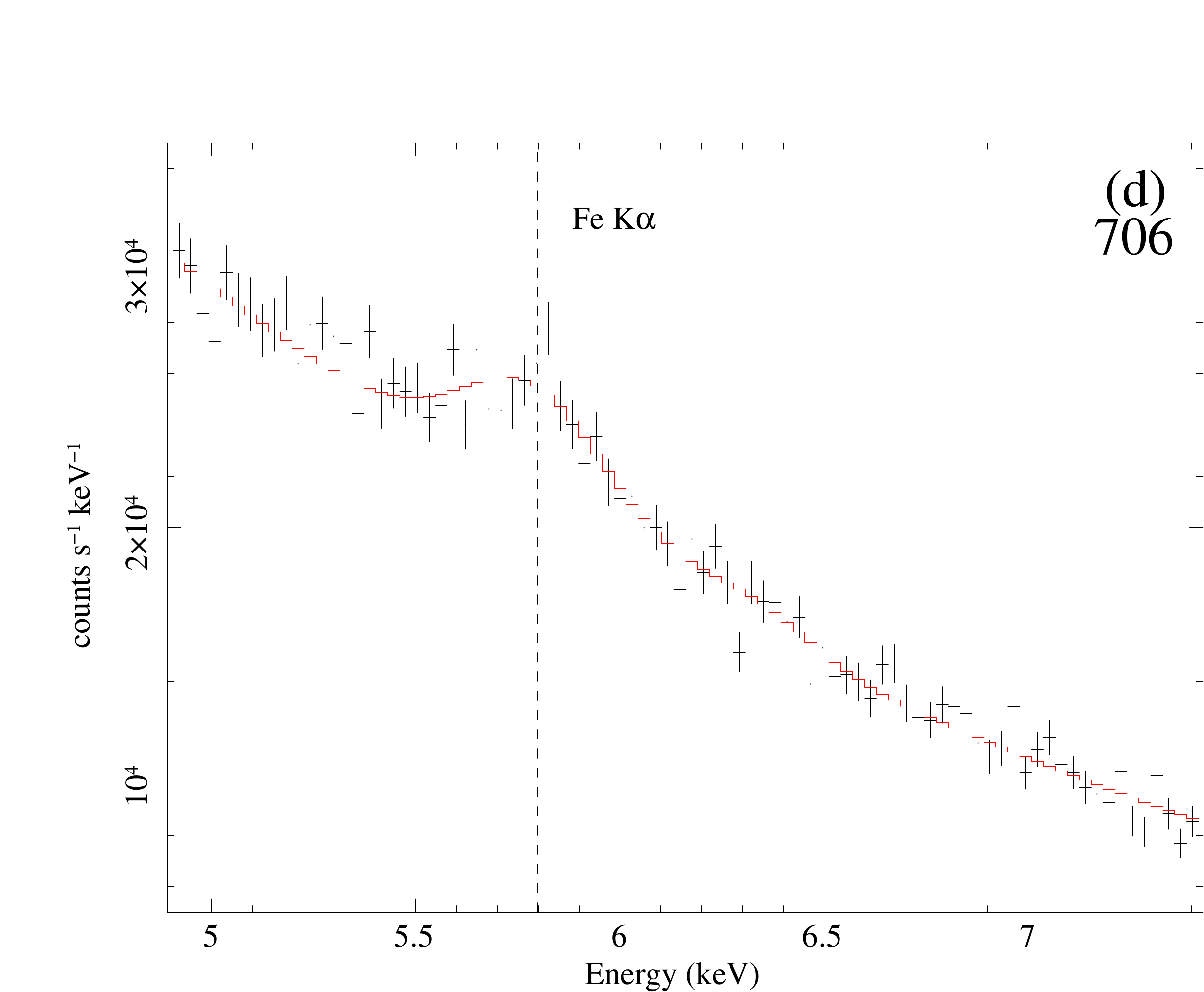}
    \end{minipage}
    \vspace{-0cm}
  \caption{\footnotesize Results of spectral fitting to \suzaku-XIS+PIN obs.~706 with \myt. Panels are as in \fr{fig-702-xis}.
  \label{fig-706-xispin}}
\end{figure*}

%% file: tab-specall.tex
\clearpage

%minipage with a vspace before its end to
%effectively center the table nicely!
\begin{minipage}{0.9\textwidth}
\begin{rotatetable}
\begin{deluxetable*}{clcccc c}
\tablecaption{\footnotesize Spectral-fitting results for \fourc\
\nustar\ and \suzaku\
observations obtained from fitting with the \myt\ model. Columns (3)
to (6) correspond to distinct fits. The contents of each row in
these columns are labeled in Column~2 as follows:
(1) Telescope and, in the case of \suzaku, obsID for each
individual fit;
(2) detector(s) associated with data for each fit;
(3) fit \cs\ and degrees of freedom (d.o.f.);
(4) reduced \cs\ (\csn) for fit;
(5) null fit probability (\pnull);
(6) angle between torus symmetry axis and observer's line-of-sight;
(7) \ccross, i.e.~either \cpx\ (\suzaku) or \cba\ (\nustar);
(8) \nhinter, which gives
the total column density of material intervening between the observer and the source; where fixed, this is the tabulated Galactic column density;
%either tabulated Galactic column density \nhgal\ (where fixed); or, otherwise, the total column density of material intervening between the observer and the source;  
(9) \nhz, i.e.~equivalent hydrogen column density associated with the single zeroth order (direct) continuum in all cases, except for \suzaku\ observation 706, where it is associated with the partially covered zeroth order continuum;
%(10) \nhztwo, only for \suzaku\ observation 706, is the equivalent neutral hydrogen column density associated with the full coverage zeroth order continuum;
(10) \nhs,
equivalent hydrogen column density associated with the scattered (reflected) continuum and the fluorescent line emission component;
(11) $A_Z$, only for \suzaku\ observation 706, is the relative normalization between fully and partially covering direct continua;
(12) $A_S$, i.e.~relative normalization between direct and scattered continuum;
(13) $\Gamma$ is the power law slope;
(14) \eshift\ is the energy shift of the \feka\ model at the line peak in the
observed frame;
(15) \feka\ line flux;
(16) \feka\ line EW;
(17) \feka\ line Gaussian width \sigl\ (see text);
(18) \feka\ line FWHM;
(19) 2--10~keV continuum flux, observed frame;
(20) 2--10~keV continuum absorbed luminosity, AGN frame;
(21) 2--10~keV continuum unabsorbed luminosity, AGN frame;
(22) 10--30~keV continuum flux, observed frame;            
(23) 10--30~keV continuum absorbed luminosity, AGN frame;  
(24) 10--30~keV continuum unabsorbed luminosity, AGN frame.
Fixed parameters are indicated by $^{\rm f}$.
%The best-fitting energy shifts
%of the \feka\ model, \eshift, are given at the line peak in the
%observed frame.  All remaining parameters are given in the AGN frame,
%except for continuum fluxes, which are in the
%observed frame, and luminosities which are in both frames.
%%\label{tab-702-706-nu}}
\label{tab-specall}}
    
  \tabletypesize{\normalsize}
  \tablehead{
    \colhead{(1)}&
    \colhead{(2)}&
    \colhead{(3)}&
    \colhead{(4)}&
    \colhead{(5)}&
    \colhead{(6)}&
    \colhead{(7)}
  }

\startdata
1 & Mission/obsID                     & \nustar\                             & \suzaku--702                   & \suzaku--702                       & \suzaku--706                          & \suzaku--706                     \\     
2 &Instruments                        & FPMA/B                               & XIS                            & XIS, PIN                           & XIS                                   & XIS, PIN                         \\
3 &\cs/d.o.f.                         & 1377.1/1211                          & 295.5/255                      & 321.6/269                          & 301.4/261                             & 310.8/273                        \\
4 &\csn\                              & 1.137                                & 1.159                          & 1.195                              & 1.155                                 & 1.138                            \\
5 &\pnull\                            & 0.001                                & 0.041                          & 0.015                              & 0.043                                 & 0.058                            \\
%6A &\thobsz\ (\degr)                  & $-$                                  & $-$                            & $-$                                & $90^{\rm f}$                           & $90^{\rm f}$                       \\   
6 &\thobs\ (\degr)                   & $0^{\rm f}$                           & $0^{\rm f}$                      & $0^{\rm f}$                          & $0^{\rm f}$                            & $0^{\rm f}$                       \\     
7 &\ccross\                           & \aer{1.03}{+0.01}{-0.01}             & $-$                            & $1.18^{\rm f}$                      & $-$                                   & \aer{1.80}{+0.30}{-0.40}          \\   
8 &\nhinter\                         & 0.116$^{\rm f}$                       & 0.116$^{\rm f}$                  & 0.116$^{\rm f}$                      & \aer{0.268}{+0.019}{-0.019}          & \aer{0.268}{+0.019}{-0.018}       \\   
9 &\nhz\ ($10^{22}$~\cunits)       & \aer{1.224}{+0.321}{-0.344}           & \aer{0.192}{+0.016}{-0.015}    & \aer{0.189}{+0.015}{-0.015}        & \aer{14.549}{+3.125}{-2.613}          & \aer{14.431}{+3.119}{-2.562}     \\
%10&\nhztwo\ ($10^{22}$~\cunits)       & $-$                                  & $-$     	              & $-$                                & 0.001$^{\rm f}$                        & 0.001$^{\rm f}$                    \\
10&\nhs\ ($10^{22}$~\cunits)          & \aer{289.960}{+52.310}{-45.530}       & \aer{40.262}{+114.648}{-23.150}& \aer{288.760}{+124.750}{-85.710}   & $>53.478$                             & \aer{149.060}{+278.200}{-97.392} \\
11&$A_Z$                              & $-$                                  & $-$	                      & $-$                                & \aer{0.325}{+0.077}{-0.087}           & \aer{0.320}{+0.085}{-0.083}      \\
12&$A_S$                              & \aer{0.716}{+0.087}{-0.083}          & \aer{0.679}{+0.535}{-0.203}    & \aer{0.702}{+0.118}{-0.114}        & \aer{1.483}{+1.155}{-0.349}           & \aer{1.404}{+0.614}{-0.310}      \\
13&$\Gamma$                           & \aer{1.919}{+0.019}{-0.020}          & \aer{1.921}{+0.015}{-0.015}    & \aer{1.914}{+0.014}{-0.015}        & \aer{2.012}{+0.038}{-0.040}           & \aer{2.012}{+0.039}{-0.040}      \\
14&\eshift~(eV)                       & \aer{0.7}{+84.2}{-51.3}              & \aer{-2.9}{+23.4}{-25.1}       & \aer{-0.5}{+23.9}{-25.2}           & \aer{16.2}{+32.5}{-35.8}              & \aer{16.2}{+32.9}{-35.1}         \\
15&\ifeka\ ($10^{-5}$~\phunits)        & \aer{1.76}{+0.21}{-0.20}             & \aer{1.92}{+1.52}{-0.57}       & \aer{1.95}{+0.33}{-0.32}           & \aer{1.97}{+1.54}{-0.47}              & \aer{1.99}{+0.87}{-0.44}         \\
16&\ewfeka\ (eV)                      & \aer{38}{+5}{-4}                     & \aer{43}{+34}{-13}             & \aer{43}{+7}{-7}                   & \aer{51}{+40}{-12}                    & \aer{51}{+22}{-11}               \\
17&\sigl\ (eV)                       & $0.850^{\rm f}$                         & $0.850^{\rm f}$                    & $0.850^{\rm f}$                       & \aer{149.4}{+42.5}{-36.1}             & \aer{149.6}{+42.6}{-36.3}        \\     
18&\fwfeka\ (\kmps)                  & $100^{\rm f}$                          & $100^{\rm f}$                   & $100^{\rm f}$                       & \aer{17574}{+5001}{-4252}             & \aer{17604}{+5018}{-4274}        \\     
19&\fcobs\ ($10^{-11}$~\funits)        & 2.87                                 & 3.15                           & 3.16                               & 2.53                                  & 2.53                             \\
20&\lcra\ ($10^{44}$~\lunits)          & 7.70                                 & 8.61                           & 8.62                               & 6.88                                  & 6.88                             \\
21&\lcru\ ($10^{44}$~\lunits)          & 10.32                                & 10.48                          & 10.53                              & 7.36                                  & 7.36                             \\
22&\fcobsH\ ($10^{-11}$~\funits)       & 2.86                                 & $-$                            & 1.14                               & $-$                                   & 0.95                             \\
23&\lcraH\ ($10^{44}$~\lunits)         & 7.94                                 & $-$                            & 9.34                               & $-$                                   & 11.46                            \\
24&\lcruH\ ($10^{44}$~\lunits)         & 7.86                                 & $-$                            & 8.07                               & $-$                                   & 4.94                             \\
\enddata                 
\end{deluxetable*}
\end{rotatetable}
\vspace{-25cm}
\end{minipage}
\clearpage

%1 
%2 
%3 
%4 
%5 
%%6
%6 
%7 
%7A
%8 
%8A
%9 
%9A
%10
%11
%12
%13
%14
%15
%15
%16
%17
%18
%19
%20
%21